%% file: 81P.tex
\documentclass[]{aastex7}
\usepackage{ulem}
\usepackage{mhchem}

\newcommand\subs[1]{\textsubscript{#1}}
\newcommand\sups[1]{\textsuperscript{#1}}
\newcommand\rh[1]{\textcolor{black}{{\textit{r}\subs{\textit{H}}}#1}}

\newcommand\trot[1]{\textcolor{black}{{\textit{T}\subs{rot}}#1}}

\newcommand\kms[1]{\textcolor{black}{{km\,s$^{-1}$}#1}}
\newcommand\ps[1]{\textcolor{black}{{s$^{-1}$}#1}}

\definecolor{gold}{rgb}{0.64,0.54,0.29}

\usepackage{color}

\newcommand{\ltsimeq}{\raisebox{-0.6ex}{$\,\stackrel
        {\raisebox{-.2ex}{$\textstyle <$}}{\sim}\,$}}
\newcommand{\gtsimeq}{\raisebox{-0.6ex}{$\,\stackrel
        {\raisebox{-.2ex}{$\textstyle >$}}{\sim}\,$}}

\usepackage{hyperref}
\usepackage{datetime2}
\usepackage{graphicx}
\usepackage{caption} 

\usepackage{amsmath}

\received{2026-08-03}
\revised{}
\accepted{}
\submitjournal{The Planetary Science Journal}

\shorttitle{A JWST Study of Stardust. I.}
\shortauthors{Roth et al.}

\begin{document}


\title{A JWST Study of Stardust. I. Infrared Spectroscopy of Comet 81P/Wild 2 and Overall Composition.}

\correspondingauthor{Nathan X. Roth}
\email{nathaniel.x.roth@nasa.gov}

\author[0000-0002-6006-9574]{Nathan X. Roth}
\affiliation{Solar System Exploration Division, Astrochemistry Laboratory Code 691, NASA Goddard Space Flight Center, 8800 Greenbelt Rd, Greenbelt, MD 20771, USA}
\affiliation{Department of Physics, American University, 4400 Massachusetts Ave NW, Washington, DC 20016, USA}
\email{nathaniel.x.roth@nasa.gov}

\author[0000-0001-7694-4129]{Stefanie N. Milam}
\affiliation{Solar System Exploration Division, Astrochemistry Laboratory Code 691, NASA Goddard Space Flight Center, 8800 Greenbelt Rd, Greenbelt, MD 20771, USA}
\email{stefanie.n.milam@nasa.gov}

\author[0000-0002-1888-7258]{Diane H. Wooden}
\affiliation{NASA Ames Research Center, Astrophysics Division, MS 245-6, Moffett Field, CA 94035-1000, USA}
\email{diane.h.wooden@nasa.gov}

\author[0000-0001-6567-627X]{Charles E. Woodward}
\affiliation{Minnesota Institute for Astrophysics, School of Physics and Astronomy, 116 Church Street SE, University of Minnesota, Minneapolis, MN 55455, USA}
\email{chickw024@gmail.com}

\author[0000-0002-8130-0974]{Dominique Bockelée-Morvan}
\affiliation{LIRA, Observatoire de Paris, Université PSL, CNRS, Sorbonne Université, Université Paris Cité, 5 place Jules Janssen, 92195 Meudon, France}
\email{dominique.bockelee@obspm.fr}

\author[0000-0002-6702-7676]{Michael S. P. Kelley}
\affiliation{Department of Astronomy, University of Maryland, College Park, MD 20742-0001, USA}
\email{msk@astro.umd.edu}

\author[0000-0001-6752-5109]{Steven B. Charnley}
\affiliation{Solar System Exploration Division, Astrochemistry Laboratory Code 691, NASA Goddard Space Flight Center, 8800 Greenbelt Rd, Greenbelt, MD 20771, USA}
\email{steven.b.charnley@nasa.gov}

\author[0000-0002-5187-1653]{Simon J. Clemett}
\affiliation{NASA Johnson Space Center, 2224 Bay Area Blvd, Houston, TX, 77058, USA}
\email{simon.j.clemett@nasa.gov}

\author[0000-0001-8233-2436]{Martin A. Cordiner}
\affiliation{Solar System Exploration Division, Astrochemistry Laboratory Code 691, NASA Goddard Space Flight Center, 8800 Greenbelt Rd, Greenbelt, MD 20771, USA}
\affiliation{Department of Physics, The Catholic University of America, 620 Michigan Ave., N.E. Washington, DC 20064, USA}
\email{martin.cordiner@nasa.gov}


\author[0000-0002-9667-5904]{Perry A. Gerakines}
\affiliation{Solar System Exploration Division, Astrochemistry Laboratory Code 691, NASA Goddard Space Flight Center, 8800 Greenbelt Rd, Greenbelt, MD 20771, USA}
\email{perry.a.gerakines@nasa.gov}

\author[0000-0001-6397-9082]{David E. Harker}
\affiliation{Department of Astronomy and Astrophysics, University of California, San Diego, 9500 Gilman Drive, MC 0424, La Jolla, CA 92093-0424, USA}
\email{dharker@ucsd.edu}


\author[0000-0002-2541-1602]{Els Peeters}
\affiliation{Department of Physics and Astronomy, University of Western Ontario, London, Ontario N6G 2V4, Canada}
\affiliation{Institute for Earth and Space Exploration, University of Western Ontario, London, Ontario N6A 5B7, Canada}
\affiliation{SETI Institute, Mountain View, CA 94043, USA}
\email{epeeters@uwo.ca}

\author[0000-0002-1883-552X]{Ella Sciamma-O’Brien}
\affiliation{NASA Ames Research Center, Astrophysics Division, MS 245-6, Moffett Field, CA 94035-1000, USA}
\email{ella.m.sciammaobrien@nasa.gov}

\author[0000-0001-6752-5109]{Geronimo L. Villanueva}
\affiliation{Solar System Exploration Division, Code 690, NASA Goddard Space Flight Center, 8800 Greenbelt Rd, Greenbelt, MD 20771, USA}
\email{geronimo.l.villanueva@nasa.gov}




\input{Abstract}



\input{Intro}

\input{Observations}
\input{Results}

\input{Conclusion}

\begin{acknowledgments}
This work is based on observations made with the NASA/ESA/CSA James Webb Space Telescope. The data were obtained from the Mikulski Archive for Space Telescopes at the Space Telescope Science Institute, which is operated by the Association of Universities for Research in Astronomy, Inc., under NASA contract NAS 5-03127 for JWST. These observations are associated with program \#1897. The specific observations analyzed can be accessed via doi:10.17909/py2d-2z26. Support for program \#1897 was provided by NASA through a grant from the Space Telescope Science Institute, which is operated by the Association of Universities for Research in Astronomy, Inc., under NASA contract NAS 5-03127. N.X.R, C.E.W, D.H.W. and M.S.P.K. acknowledge support from JWST-GO-01897. N.X.R., M.A.C., S.B.C., S.N.M., and P.A.G. were also supported by the NASA Planetary Science Division Internal Scientist Funding Model program through the Fundamental Laboratory Research work package (FLaRe). D.H.W. was also supported by the NASA Planetary Science Division Internal Scientist Funding Model through the Cold Solar System Objects (CSSO) ISFM. We gratefully acknowledge insightful discussions with Keiko Nakamura-Messenger, Don E. Brownlee, John P. Bradley, and Hope A. Ishii regarding the Stardust returned samples.
\end{acknowledgments}

\software{Astropy \citep{astropy:2013, astropy:2018, astropy:2022},
Astroquery \citep{Ginsburg2019}, jwstComet \citep{Roth2026b},
sbpy \citep{Mommert2019},
SciPy \citep{Virtanen2020}}

\input{ThermalAppendix}


\clearpage

\bibliography{81P}{}
\bibliographystyle{aasjournalv7}



\end{document}

%% file: Abstract.tex
\begin{abstract}

We report observations of comet 81P/Wild 2, target of the Stardust sample return mission, on UT 2023 March 20 and 24 at a heliocentric distance (\rh{}) of 1.85 au using the NIRSpec and MIRI integral field unit spectrographs on board the James Webb Space Telescope (JWST). This  study is the first compositional comparison between JWST remote-sensing spectroscopy of a solar system object against terrestrial analysis of its returned samples. We securely detected molecular emission from \ce{H2O}, \ce{CH4}, \ce{C2H6}, \ce{CH3OH}, CO, \ce{CO2}, \ce{^13CO2}, OCS, HCN, and CN and find molecular abundances  consistent within $2\sigma$ with those reported during previous perihelion passages. The water ortho-to-para ratio was $2.76\pm0.05$, and the $\ce{^12CO2}/\ce{^13CO2}$ ratio was $85\pm4$. Thermal emission from the nucleus and dust was detected and modeled, providing an effective nucleus radius of $1.77\pm0.04$ km and a dust composition (relative mass fraction of the submicron grains) of $\sim36\%$ amorphous carbon, $\sim25\%$ amorphous Mg:Fe olivine, $\sim23\%$ Mg-rich crystalline olivine, and $\sim15\%$ amorphous Mg:Fe pyroxene. The crystalline mass fraction of the sub-micron grains in the coma was $0.362\pm0.003$. Comparison of the JWST-derived thermal model against the fine-grained materials in Stardust returned samples demonstrates complementarity between the missions, with each most sensitive to a different population of the coma dust grains. 

\end{abstract}

%% file: Intro.tex
\section{Introduction} \label{sec:intro}
Comets may serve as ``living fossils'' of solar system formation, with the volatile composition of their nuclei reflecting the chemistry and prevailing conditions present where and when they formed \citep{Bockelee2004,Mumma2011a}. The James Webb Space Telescope \cite[JWST;][]{2023PASP..135d8001R,gardner23-jwst} is opening new frontiers in cometary science, providing sensitive measures of the volatile and refractory components of the coma, including evidence for polycyclic aromatic hydrocarbon (PAH) species \citep{2025PSJ.....6..139W}. Here we report JWST MIRI and NIRSpec integral field unit (IFU) observations of Jupiter-family comet 81P/Wild 2 (hereafter 81P). Comet 81P represented an exceptional JWST target, being also the  target of the Stardust sample return mission \citep[e.g.,][]{Sandford2006,2010M&PS...45..701C,Brownlee2017} and complemented by the availability of ground-based, high spectral resolution near-infrared compositional studies during its previous perihelion passages \citep{DelloRusso2014}. These combined measurements reported here represent the first opportunity to integrate the results of comet sample return analyses with a full near--mid infrared cometary spectrum measured with the unparalleled sensitivity of JWST.

This manuscript (Paper I) reports our radiative transfer modeling of molecular emission from volatiles in the coma and thermal modeling of continuum emission from the nucleus and coma dust grains. Our best-fit models for all three of these quantities were subtracted to form a residual spectrum, which is analyzed in Paper II \citep{Roth2026c} to test for the presence of polycyclic aromatic hydrocarbons (PAHs). Section~\ref{sec:obs} details our observations and data reduction procedures. Section~\ref{subsec:nucleus} provides our modeling of the nucleus signal. Sections~\ref{subsec:h2o} and \ref{subsec:nirspec} describe our analysis of molecular emission from water and trace volatiles in the coma. Section~\ref{subsec:cfx-dtm} shows our thermal modeling formalism for the dust signal. Section~\ref{sec:dw-dust-bestfit} discusses the results of our best-fit dust model, and Section~\ref{sec:dw-dust-stardust} compares the JWST model to the dust properties found in the Stardust returned samples.

%% file: Observations.tex
\section{Observations and Data Reduction} \label{sec:obs}
We conducted post-perihelion observations toward 81P on UT 2023 March 20 and 24 using JWST. An observing log is in Table~\ref{tab:obslog}. We used the the MIRI Medium Resolution Spectroscopy IFU on 20 March \citep{Wright2023}, covering wavelengths from $4.9 - 28.1$ $\mu$m with $\lambda/\Delta\lambda\sim3000$, and the NIRSpec IFU \citep{Boker2022} with the G395H/F290LP grating on 24 March, covering wavelengths from $2.87 - 5.27$ $\mu$m with a resolving power $\lambda/\Delta\lambda\sim2700$. Observations were conducted without target acquisition (following Cycle 1 best practices for faint, extended objects such as 81P) using JPL/HORIZONS ephemerides (JPL\#K222/37). We used a four-point dither pattern, with dedicated background exposures taken offset 180$\arcsec$ from the comet ephemeris position along the sunward position angle.

\begin{deluxetable*}{ccccccccc}[h]
\tablenum{1}
\tablecaption{Observing Log\label{tab:obslog}}
\tablewidth{0pt}
\tablehead{
\colhead{JWST} & \colhead{IFU} & \colhead{No. Dithers} & \colhead{UTC Midpoint} & \colhead{$t_{int}$} & \colhead{\rh{}} & \colhead{$\Delta$} & \colhead{$\phi_{\mathrm{STO}}$} & \colhead{$\psi_{\sun}$} \\
\colhead{Archive No.} & \colhead{Configuration} & \colhead{} & \colhead{} & \colhead{(s)} & \colhead{(au)} & \colhead{(au)} & \colhead{($\degr$)} & \colhead{($\degr$)}
}
\startdata
\multicolumn{9}{c}{MIRI IFU} \\
\hline
3 (comet) & MRS\_LONG\_C & 4 & 2023 Mar. 20 21:52:18 & 166 &  1.87 & 1.43 & 32.6 & 272.7 \\ 
4 (bkgnd) & MRS\_LONG\_C & 4 & & & & & & \\
3 (comet) & MRS\_MEDIUM\_B & 4 & 2023 Mar. 20 22:12:05 & 277 &  1.87 & 1.43 & 32.6 & 272.7 \\ 
4 (bkgnd) & MRS\_MEDIUM\_B & 4 & & & & & & \\
3 (comet) & MRS\_SHORT\_A & 4 & 2023 Mar. 20 22:35:20 & 666 &  1.87 & 1.43 & 32.6 & 272.7 \\ 
4 (bkgnd) & MRS\_SHORT\_A & 4 & & & & & & \\
\hline
\multicolumn{9}{c}{NIRSpec IFU} \\
\hline
1 (comet) & NIRSpec G395H & 4 & 2023 Mar. 24 07:41:10 & 2567 & 1.85 & 1.42 & 32.2 & 272.3 \\
2 (bkgnd) & NIRSpec G395H & 4 &  & & &  &  & 
\enddata
\tablecomments{\textit{t}\subs{int} is the total on-source integration time. \rh{}, $\Delta$, $\phi_\mathrm{STO}$, and $\psi_{\sun}$ are the heliocentric distance, geocentric distance,
phase angle (Sun--Comet--Earth), and position angle of the Sun--Comet radius vector measured counter-clockwise (east) from celestial north, respectively, of 81P at the time of observations.}
\end{deluxetable*}

We followed data reduction and analysis procedures as detailed by \cite{2025PSJ.....6..139W} and \citet{Roth2026a} using the \texttt{jwstComet} package \citep{Roth2026b}. Here we will only describe aspects of our reduction and analysis that differed significantly. We downloaded the Level 0 data files for NIRSpec and MIRI from the Mikulski Archive for Space Telescopes (MAST) and reduced locally using version 1.14.0 of the JWST pipeline and CRDS file version 1228. We utilized an MRS defringing algorithm in Stage 1 (calwebb2), and background subtraction in Stages 2 and 3 for NIRSpec and MIRI, respectively. The resulting Level 3 background-subtracted three-dimensional spatial-spectra data cubes (s3d) were used for analysis in this study.

\begin{figure}
\includegraphics[scale=0.47]{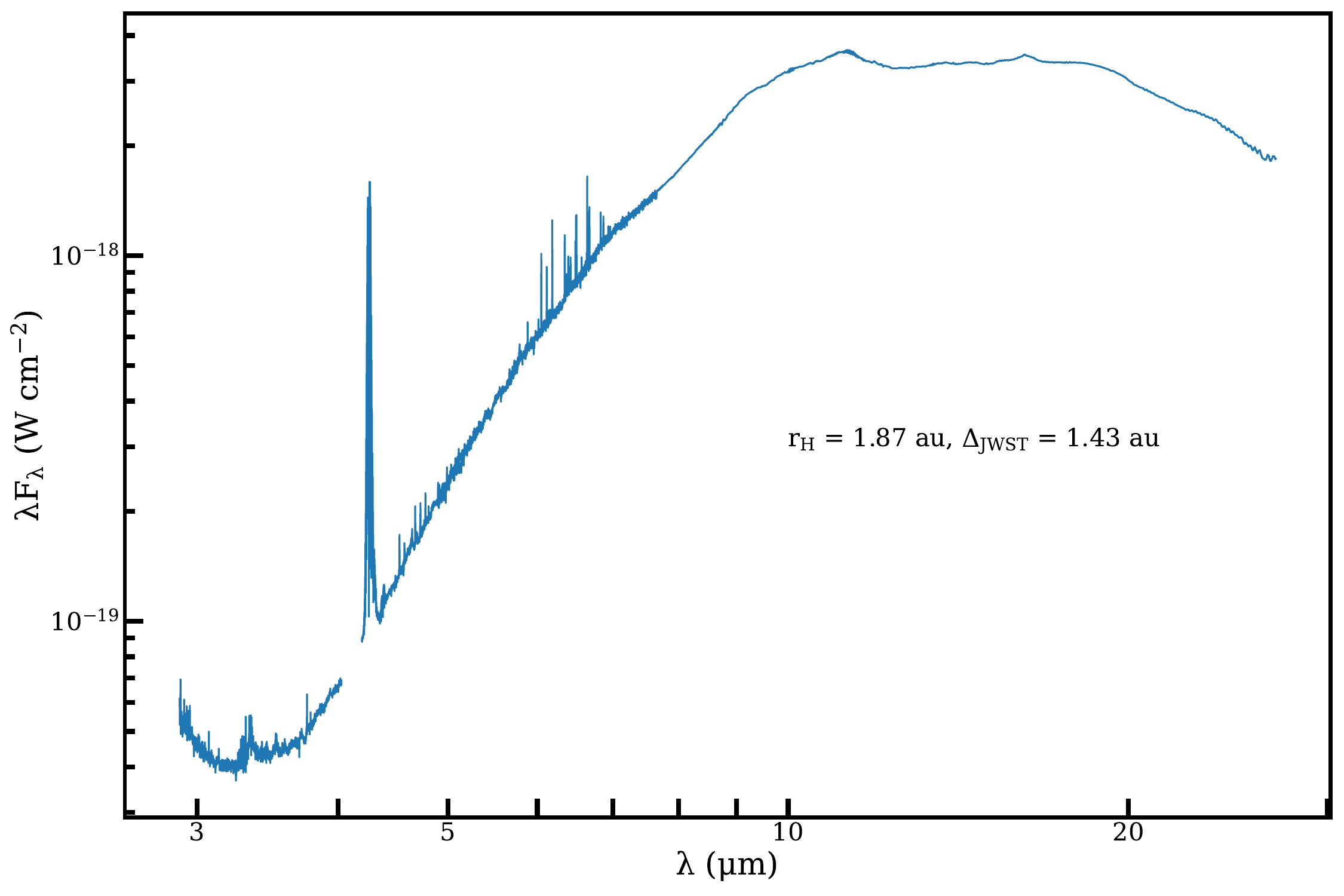}
\caption{Composite 81P/Wild 2 JWST spectrum extracted from a $1\farcs41$ diameter circular aperture centered on the comet photocenter. Wavelengths $>$ 26.0~\micron{} are clipped out due to poor MRS performance. \label{fig:complete_spectrum}}
\end{figure}

We identified spatially extended emission from both the gas and dust in the NIRSpec and MIRI data cubes. We used the Python \texttt{photutils} package  to extract spectra in 1\farcs41 diameter apertures centered on the comet photocenter. Note that in this aperture NIRSpec contains all of the nucleus flux but MIRI does not, thereby requiring application of an Aperture Correction Factor (ACF) to assess the nucleus flux in the aperture (see \S~\ref{subsubsec:dw-dust-nirspec}). We investigated similar extracts at off-nucleus positions, but none displayed a sufficient signal-to-noise ratio (S/N) for a well-constrained analysis aside from $^{12}$CO$_2$ in NIRSpec and H$_2$O in MIRI. Thus, we performed spaxel-by-spaxel analysis (a spaxel, or ``spectral pixel'' refers to the fact that each spatial element of the IFU contains a spectrum of the source at that position) of these two molecules and calculated bulk abundances for the remaining composition using our nucleus-centered extracts only. The complete 81P JWST spectrum is shown in Figure~\ref{fig:complete_spectrum}. No scaling was applied to account for potential coma or nucleus signal variability with time between the NIRSpec and MIRI observations. 

%% file: Results.tex
\input{results-nucleus}
\input{results-H2O}
\input{results-volatiles}

\input{results-thermal}

%% file: results-nucleus.tex
\section{Nucleus Extraction and Model}\label{subsec:nucleus}
Thermal emission from cometary nuclei can contribute a substantial fraction of the total emission in a spectrum, which, if not addressed, can bias a retrieval of the composition \citep{Kelley2009}.  The Stardust spacecraft imaged the resolved nucleus of comet 81P, but due to the slow rotation and fast flyby only a partial shape model could be constructed.  Furthermore, the spacecraft did not carry an infrared instrument, so the thermophysical state of the nucleus was not assessed.  Therefore, we measured the nucleus emission directly from the MIRI MRS data by separating the point source nucleus from the extended coma with a morphological technique \citep{Lamy1996,Hui2018}.  This technique uses the surface brightness distribution of the coma to predict its contribution to the inner, spatially unresolved region around the nucleus.  A model comet surface brightness distribution, $I_C$, is produced from the sum of a model coma and model nucleus,
\begin{equation}
    I_C = (I_c \rho^k + F_N \delta) * I_{PSF},
\end{equation}
where $I_c \rho^k$ is the coma surface brightness distribution, $F_N$ is the total nucleus brightness, $\delta$ is the Dirac delta function, $*$ is the convolution operator, and $I_{PSF}$ is the point spread function for the instrument and wavelength range being tested.  The coma is described by a power-law distribution as a function of projected distance to the nucleus, $\rho$, with a slope $k$.

We first generated images of 81P averaging over three MRS wavelength regions: 5.8--6.0~\micron, 6.57--6.62~\micron, and 12.5--12.7~\micron.  The wavelength intervals were chosen to avoid strong features from gas, i.e., the 6~\micron{} $\nu_2$ H$_2$O band, and the silicate emission bands at $\sim10$ and $\sim20$~\micron.  Silicate emission features from the coma can be stronger than those from the nucleus \citep{Kelley2017}.  Avoiding the band would therefore increase the nucleus-to-coma contrast.  For each wavelength range, we averaged together monochromatic Point Spread Functions (PSF) generated at each wavelength step in the MRS data cubes, producing three final PSF files.

Each image was azimuthally averaged, centered on an inner 3$\times$3 or 5$\times$5 pixel box.  Linear least-squares techniques (\texttt{scipy}'s nnls and leastsq \citet{Virtanen2020}) were used to derive the slope, coma, and nucleus scale factors.  However, instead of minimizing $\chi^2=\sum_i(I_{C,i} - I_{obs,i})^2/\sigma^2_{obs,i}$, we weighted the difference between the model and data by $\rho$: $(\chi^\prime)^2=\sum_i(I_{C,i} - I_{obs,i})^2\rho_i^2/\sigma^2_{obs,i}$.  Without the radial weights, the best-fits were more strongly driven by the PSF core and the power-law behavior of the surrounding coma was not as well reproduced.  The best-fit coma slopes were --1.40, --1.32, and --1.12 for the 5.90, 6.60, and 12.60~\micron{} images, respectively.  The best-fit nucleus brightnesses were $6.64\pm0.07$~mJy, $11.70\pm0.05$~mJy, and $61.37\pm0.03$~mJy, respectively.  Following \citet{Law2024}, we added a 2\% calibration error for final uncertainties of 0.15, 0.24, and 1.23~mJy.

The observed nucleus brightnesses were then modeled with the Near-Earth Asteroid Thermal Model (NEATM; \citealt{Harris1998}), a model that is commonly used to measure the sizes of comets from thermal emission photometry \citep{Fernandez2013}.  Briefly, the model assumes the nucleus is a sphere in instantaneous local thermodynamic equilibrium (LTE) with sunlight.  The temperature of a surface element on the nucleus is described by
\begin{equation}
T(\theta_z) = 
\left\{
  \begin{array}{lr}
   \left( \frac{(1 - A) F_\odot \cos(\theta_z)}{\eta \epsilon \sigma_{SB}} \right)^{1/4} & \textrm{for }\theta_z < 90\degr,\\
    0 & \textrm{for }\theta_z \geq 90\degr,
  \end{array}
\right.
\end{equation}
where $A=0.02$ is the assumed bolometric albedo, $F_\odot$ is the flux (irradiance) of the Sun at the heliocentric distance of the comet (1367.6 W~m$^{-2}$ at 1~au), $\theta_z$ is the zenith angle of the Sun, $\eta$ is the infrared beaming parameter, $\epsilon=0.95$ is the infrared emissivity, and $\sigma_{SB}$ is the Stefan-Boltzmann constant.  The parameter $\eta$ adjusts the effective temperature to account for surface roughness and thermal inertia (i.e., non-instantaneous equilibrium).  Optimizing the model with the observed nucleus brightness and a least-squares fit approach results in a radius of $1.77\pm0.04$~km, and a beaming parameter of $0.77\pm0.02$.  The result suggests we observed the nucleus near the minimum cross-sectional area.  The triaxial ellipsoid model of \citet{Duxbury2004} has dimensions $1.65\times2.00\times2.75$~km, and a minimum effective radius of 1.82~km.

%% file: results-H2O.tex
\section{Analysis of H$_2$O Emission}\label{subsec:h2o}

\begin{figure}
    \includegraphics[scale=0.5]{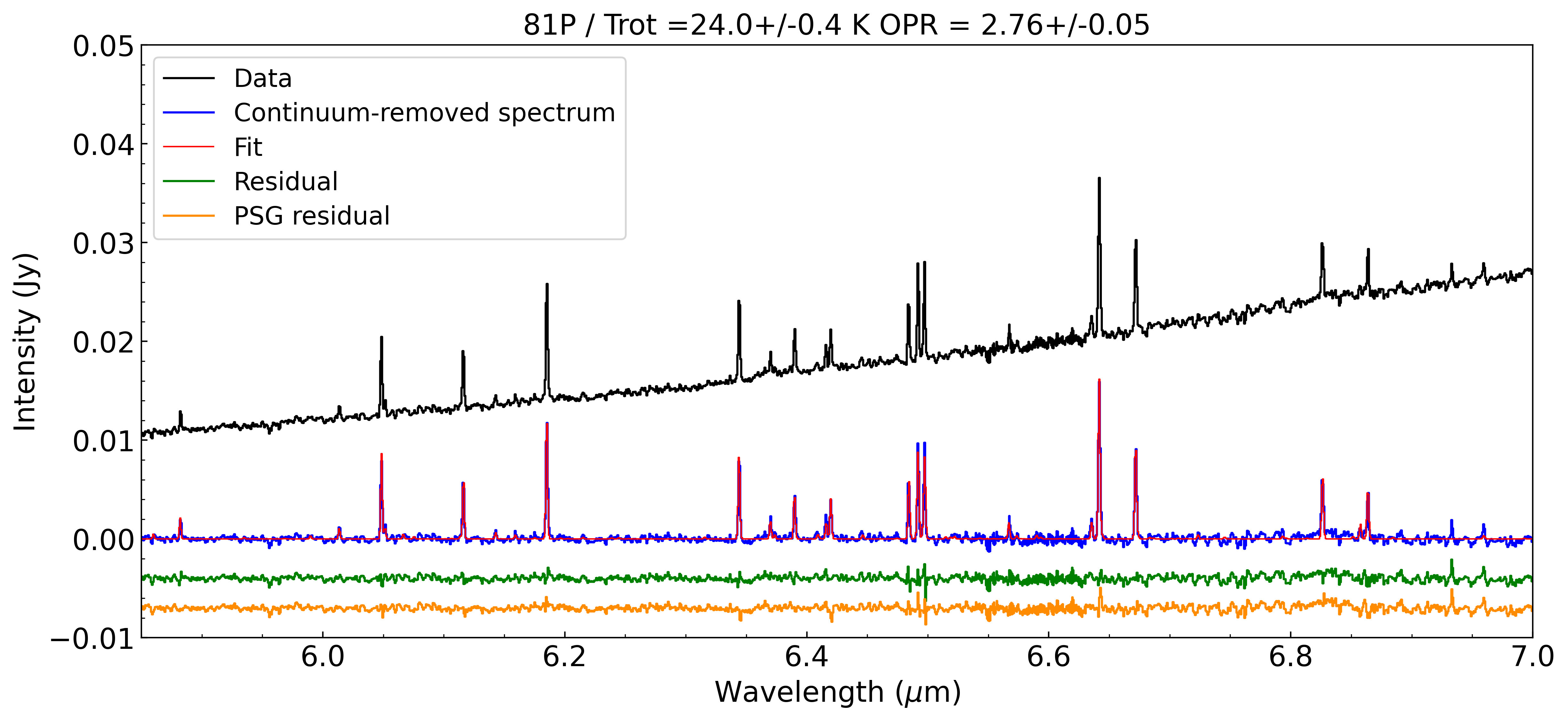}
\caption{Model fits to H$_2$O MIRI/MRS spectrum. Black line:  observed spectrum over a 1.41 arcsec diameter aperture. Blue line: continuum-subtracted spectrum. Red line: model fit using the fluorescence model of \citet{2009diwo.conf..249C}. Residuals are shown in green and retrieved parameters are given in the figure title. The residuals in orange color are those obtained using PSG retrieval tool, with inferred parameters: $Q$(H$_2$O) = 1.80 $\pm$0.04 10$^{27}$ s$^{-1}$ and $T_{\rm rot}$ = 26.6$\pm$0.4 K \label{fig:H2O-fit}}
\end{figure}

Emission lines from H$_2$O are detected in both NIRSpec and MIRI/MRS spectra of 81P. In the 4.5--5.2 $\mu$m range, faint H$_2$O ro-vibrational lines are from the $\nu_3$-$\nu_2$ and $\nu_1$-$\nu_2$ hot bands. As shown in Fig.~\ref{fig:H2O-fit}, the spectral region 5.5--7.25 $\mu$m, covered by Channel 1 of MIRI/MRS, shows strong lines from the $\nu_2$ fundamental vibrational band, with some minor contribution expected from hot bands (e.g., $\nu_3+\nu_2-\nu_3$, 2$\nu_2-\nu_2$). 

The spectral resolution of MIRI/MRS is appropriate to investigate the ortho-to-para ratio (OPR) of water molecules from the $\nu_2$ ro-vibrational lines. For the OPR analysis, we used the H$_2$O fluorescence model of \citet{2009diwo.conf..249C}, which includes
fundamental bands and hot bands and uses the comprehensive H$_2$O ab initio database of \citet{2000JChPh.113.6592S}. One should note that the 6.3-$\mu$m synthetic spectra from \citet{2009diwo.conf..249C} closely resemble those obtained by the NASA Planetary Spectrum Generator \citep[PSG;][]{Villanueva2018}, which uses the BT2 ab initio database of \citet{2006PhDT.......350B,Villanueva2012b}. 

The fitting procedure consisted of a simultaneous fit to the H$_2$O lines and the underlying continuum, represented as a cubic spline. The procedure corrects for small deviations in the frequency calibration. The water spectrum was fitted with the above-mentioned optically thin fluorescence model parametrized by a normalizing factor (proportional to the water production rate $Q$(H$_2$O)), the rotational temperature $T_{\rm rot}$ in the ground vibrational state, and the ortho-to-para ratio (OPR). Minimization was done with the Levenberg–Marquardt algorithm. The same approach was used for the analysis of C/2017 K2 MIRI/MRS data \citep{2025PSJ.....6..139W}. A fit of synthetic spectra produced by the PSG show that the OPR derived using the PSG H$_2$O fluorescence model (version 2026 March 23) would have been almost identical. Figure~\ref{fig:H2O-fit} shows the 5.5--7.25 $\mu$m MIRI spectrum extracted over a 1.4 arcsec-diameter circular aperture centered on the nucleus, with the model fit superimposed. The retrieved model parameters are $T_{\rm rot}$ = 24$\pm$0.4 K and OPR = 2.76 $\pm$ 0.05. 

In Fig.~\ref{fig:H2O-fit} the difference between the measured and modeled H$_2$O spectra (residuals) is shown in green. Also shown in orange are the residuals obtained using the PSG retrieval tool, used to determine $Q$(H$_2$O). One should note that, in PSG, the OPR is not a free parameter. Instead, $g-$factors for ortho- and para-\ce{H2O} are generated assuming the statistical equilibrium value of 3, so that $\mathrm{OPR}= 3 N(o-\ce{H2O}) / N(p-\ce{H2O})$ as modeled with the PSG \citep[see ][ for more details]{Villanueva2025}. Residuals show that neither fluorescence model is able to correctly fit the H$_2$O lines near 6.5 $\mu$m. The strength of the lines at 6.933 $\mu$m and 6.96 $\mu$m are also not explained, the former line being unidentified \citep[and not present in C/2017 K2 MIRI spectrum;][]{2025PSJ.....6..139W} and the latter corresponding to the H$_2$O $\nu_2$ $3_{21}$--$4_{32}$ ro-vibrational line. These discrepancies possibly originate from non-LTE populations in the ground vibrational state. For this weakly active comet, non-LTE populations can be expected within the field-of-view ($\sim$ 730 km radius) because the gas density is not high enough to maintain thermal equilibrium through collisions \citep[see H$_2$O rotational populations for low production rates from][]{Lee2011}. Since the higher energy levels start deviating
from the thermal equilibrium population at smaller cometocentric distances than the lower energy levels, higher discrepancies might be expected for the low-intensity infrared lines. Focusing on the eight strongest and non-blended ortho lines, we found that the ratios of observed fluxes and modeled fluxes $F$/$F_{mod}$, normalized to their mean value ($<$$F$/$F_{mod}$$>$), show a minimum standard deviation for $T_{\rm rot}$ = 24 K, with a maximum deviation for individual lines of 15\%, suggesting that LTE is a good approximation for these lines. A similar result is obtained for the five strongest para lines. Fixing $T_{\rm rot}$ to 24 K and using these eight ortho and five para lines, minimization of the standard deviation of the $F$/$F_{mod}$$/$$<$$F$/$F_{mod}$$>$ values is obtained for OPR = 2.70 $\pm$ 0.09, which is consistent with the value obtained by fitting the full spectrum.

For the determination of the water production rate from lines covered by NIRSpec and MIRI, we used the PSG retrieval tool \citep{Villanueva2018} for consistency with other production rate determinations. With the assumed expansion velocity of 0.55 km/s following the empirical relationship $v=0.8/\sqrt{r_H}$ \kms{} \citep{Biver1999,Ootsubo2012}, the derived value from the 5.5--7.25 $\mu$m MIRI spectrum extracted over a 1.4 arcsec-diameter is $Q$(H$_2$O) = $1.80 \pm0.04\times10^{27}$ s$^{-1}$. 

%% file: results-volatiles.tex
\section{Analysis of Trace Volatile Emission}\label{subsec:nirspec}
We securely detected molecular emission from multiple trace species, including \ce{CO2}, \ce{^13CO2}, OCS, \ce{CH4}, \ce{CH3OH}, \ce{C2H6}, HCN, and CN, and determined stringent $(3\sigma)$ upper limits on the production of \ce{H2CO}. \ce{CO2} was detected through both its 15 $\mu$m and 4.25 $\mu$m bands with MIRI and NIRSpec, respectively. We determined contributions from continuum and gaseous emissions using the Optimal Estimation Method implemented in the NASA PSG. We subtracted continuum thermal emission from the nucleus and dust coma, as well as scattered sunlight and instrumental artifacts, using a low-order polynomial ($3-7$, depending on the shape of the underlying continuum across the width of the spectral extract; higher order polynomials were used for the widest extracts) spectral baseline. We chose the lowest order possible to reproduce the spectral shape and avoided higher-order polynomials to prevent introducing artifacts into the spectra. The spectral baseline was fit simultaneously with the molecular emission models, thereby propagating uncertainties on the baseline fit to each retrieved quantity ($Q$ or \trot{}). The fits included a correction for opacity effects for each species  \citep[see ][for further details]{Villanueva2025,Roth2023}. Uncertainties on the derived parameters were retrieved from the diagonal elements of the covariance matrix, scaled by the square root of the reduced $\chi^2$ statistic of the fit. 

\begin{figure*}
\gridline{\fig{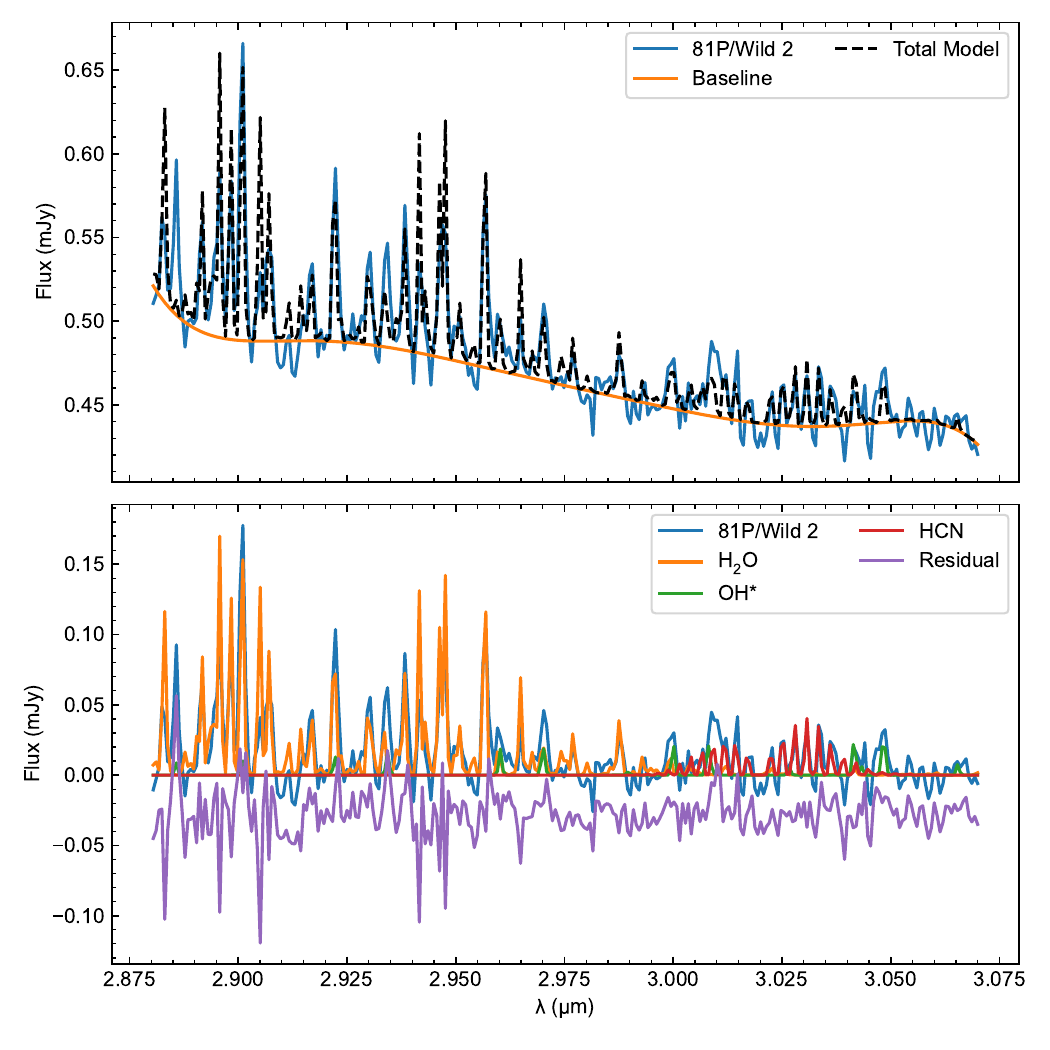}{0.45\textwidth}{(A)}
          \fig{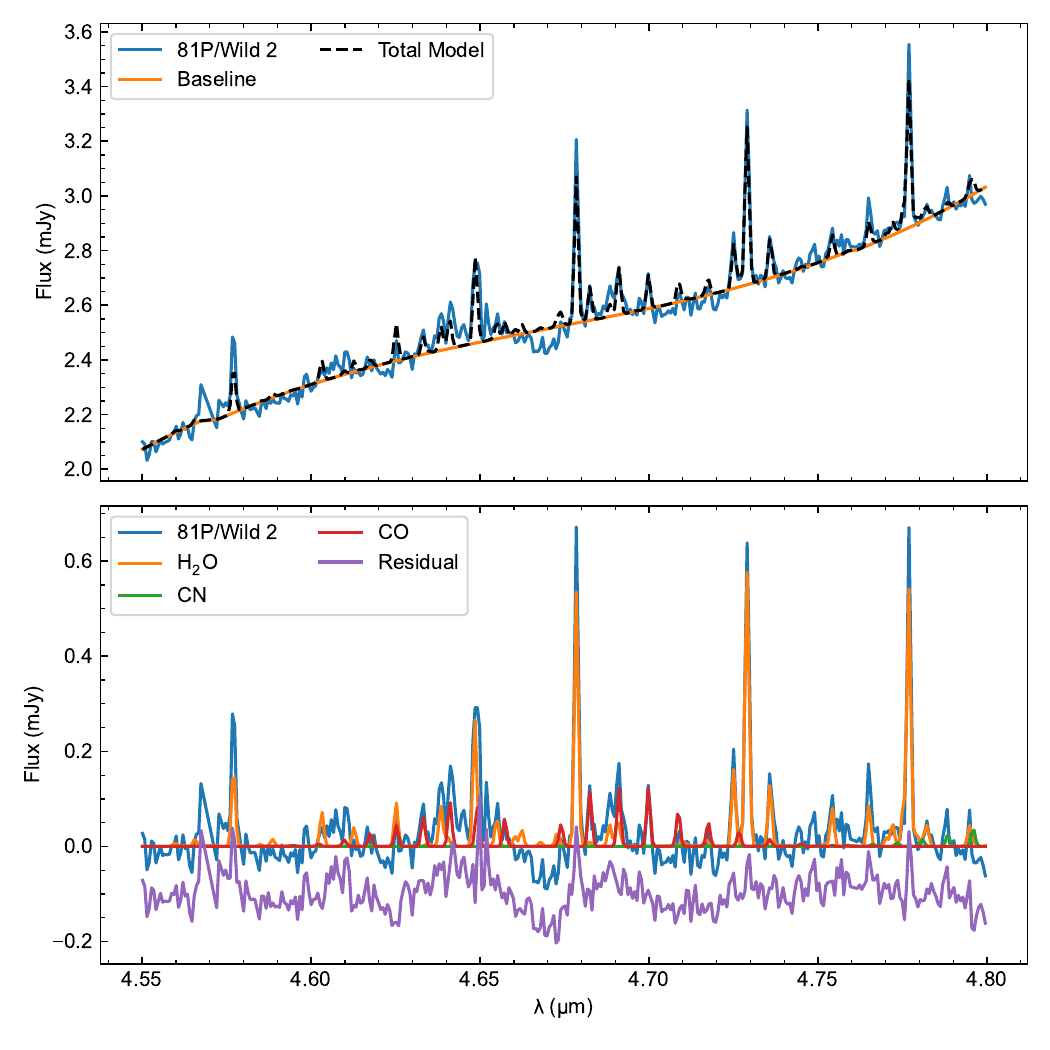}{0.45\textwidth}{(B)}
}
\gridline{\fig{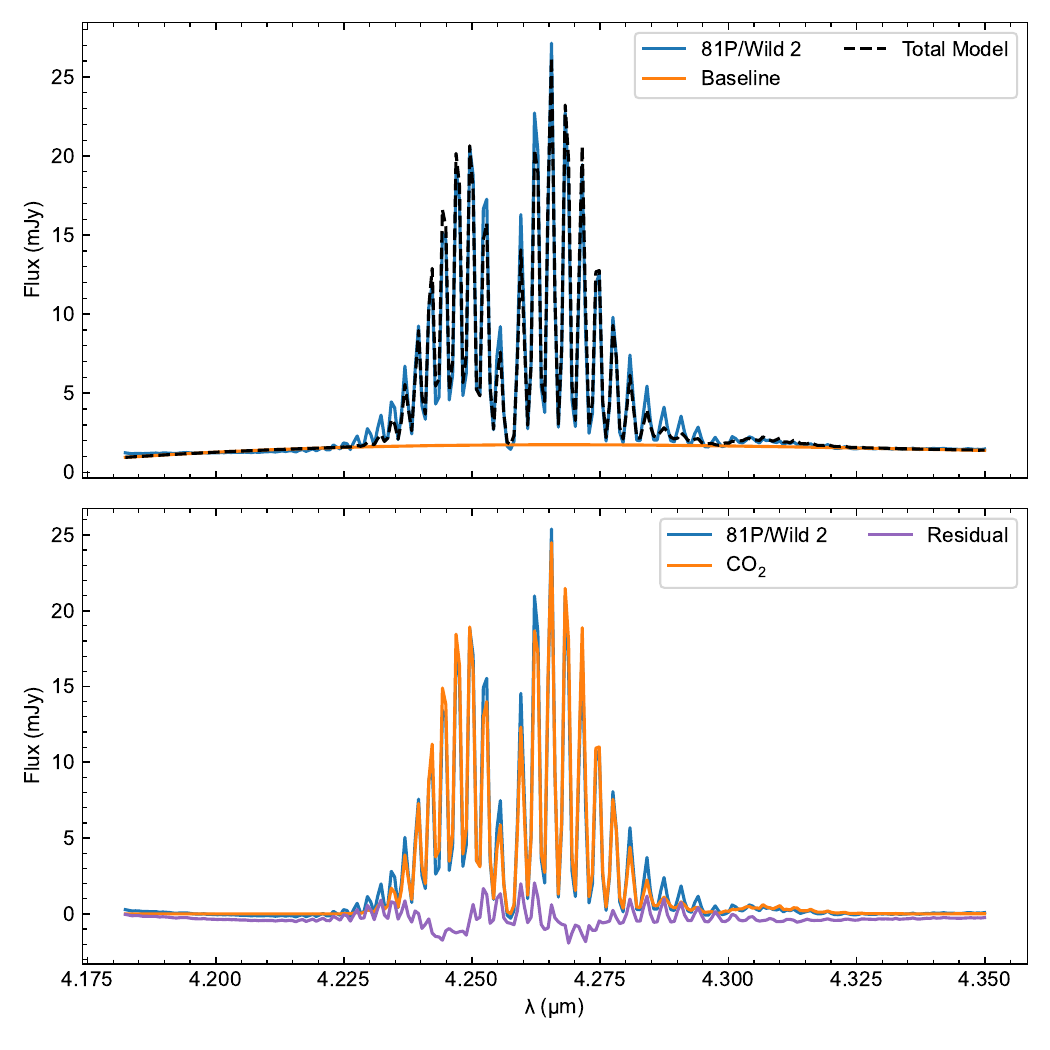}{0.45\textwidth}{(C)}
          \fig{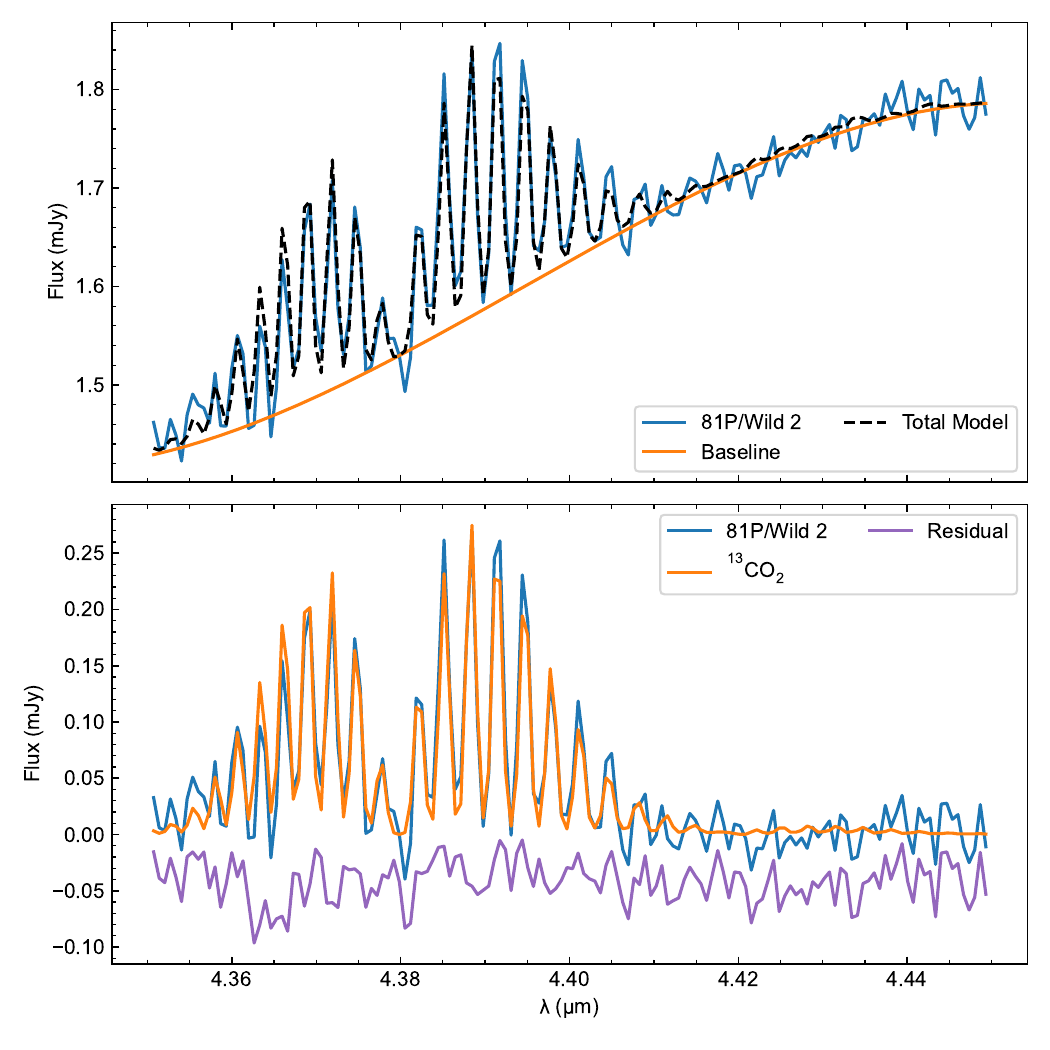}{0.45\textwidth}{(D)}
}
\caption{\textbf{(A)--(D).}  Spectra of comet 81P extracted in a 1\farcs41 diameter aperture and showing detections of \ce{H2O}, HCN, CO, CN, \ce{CO2}, and \ce{^13CO2} in 81P on March 24. For each figure, the upper panel shows the observed spectrum, spectral baseline, and total fluorescence model. The lower panel shows the baseline-subtracted spectrum and individual best-fit molecular fluorescence models. The residual spectrum (observed$-$baseline$-$models) is shown for comparison and offset vertically.
\label{fig:spec-panels1}}
\end{figure*}

\begin{figure*}
\gridline{\fig{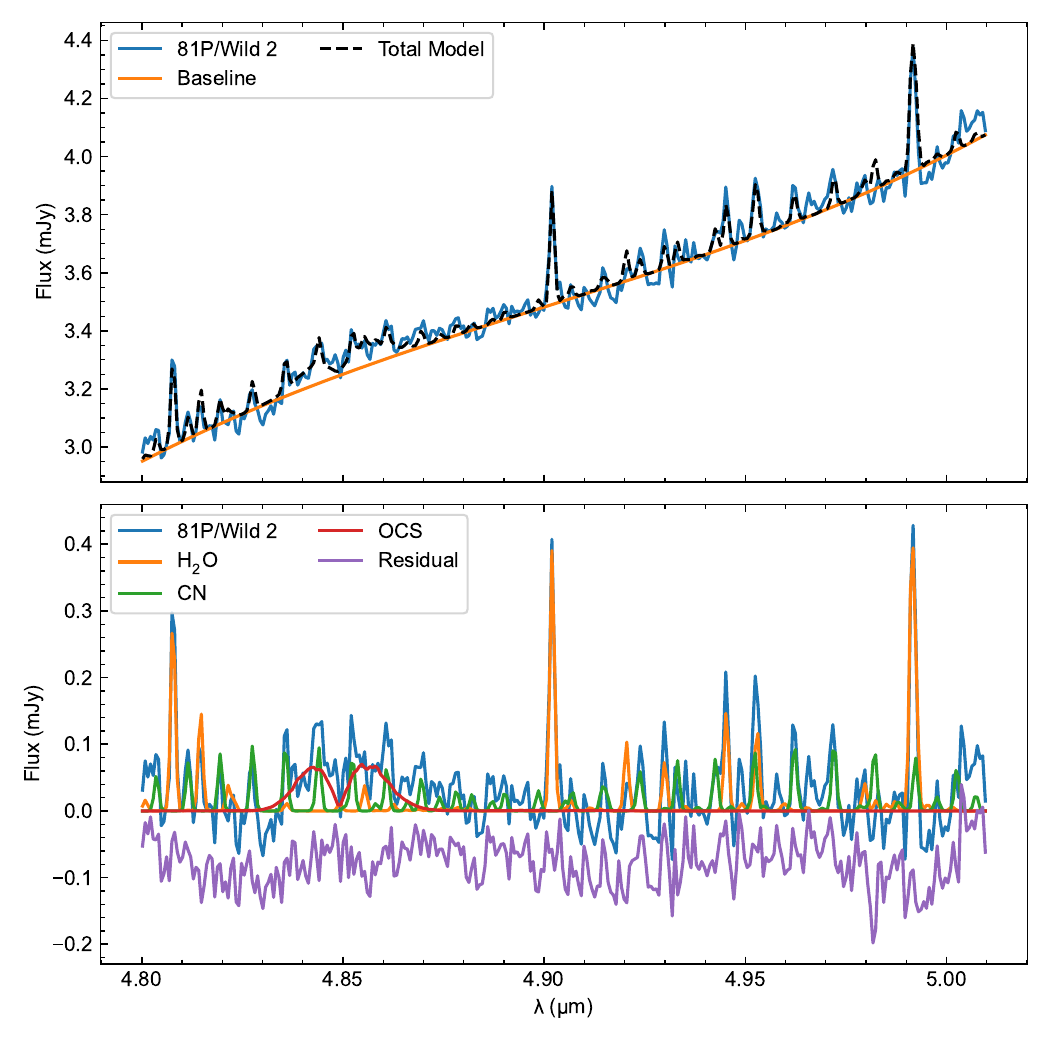}{0.45\textwidth}{(A)}
          \fig{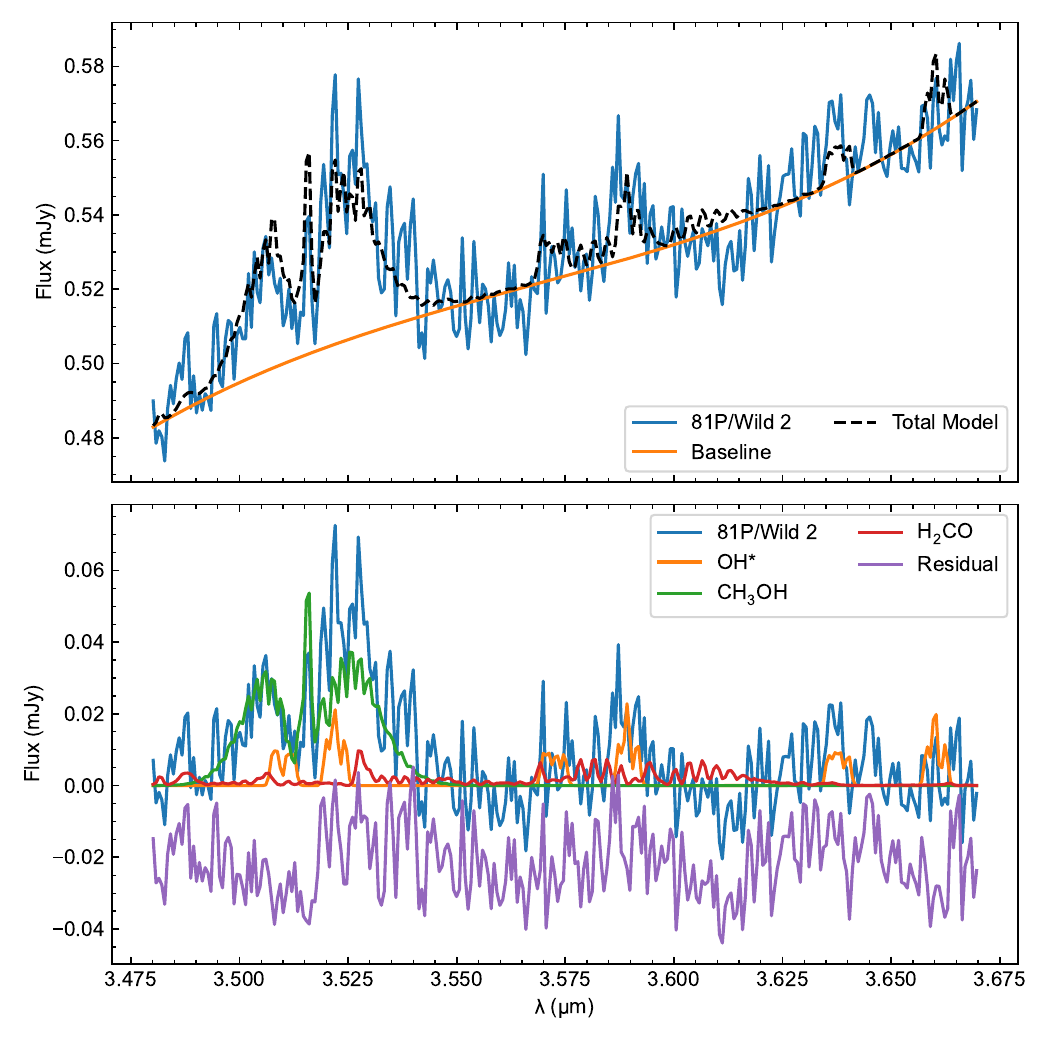}{0.45\textwidth}{(B)}
}
\gridline{\fig{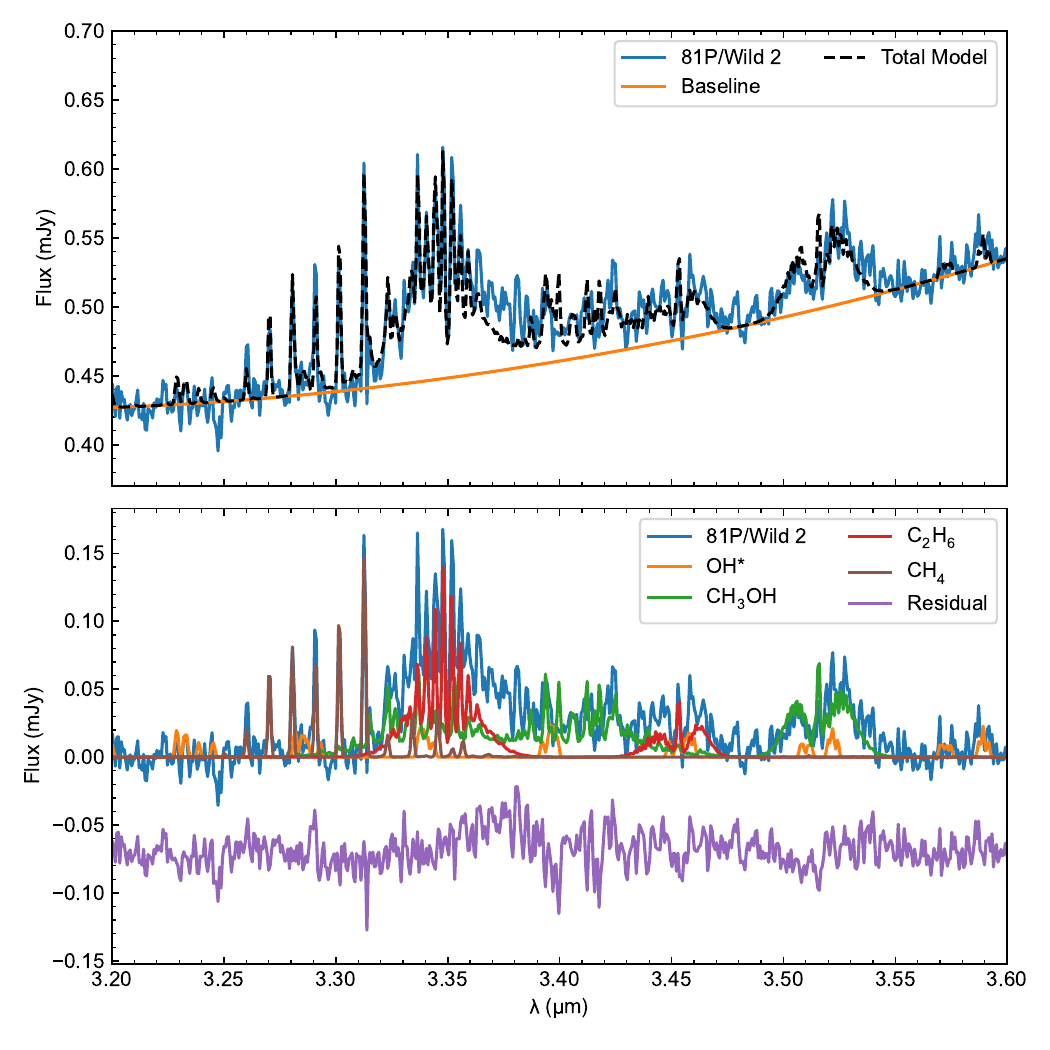}{0.45\textwidth}{(C)}
          \fig{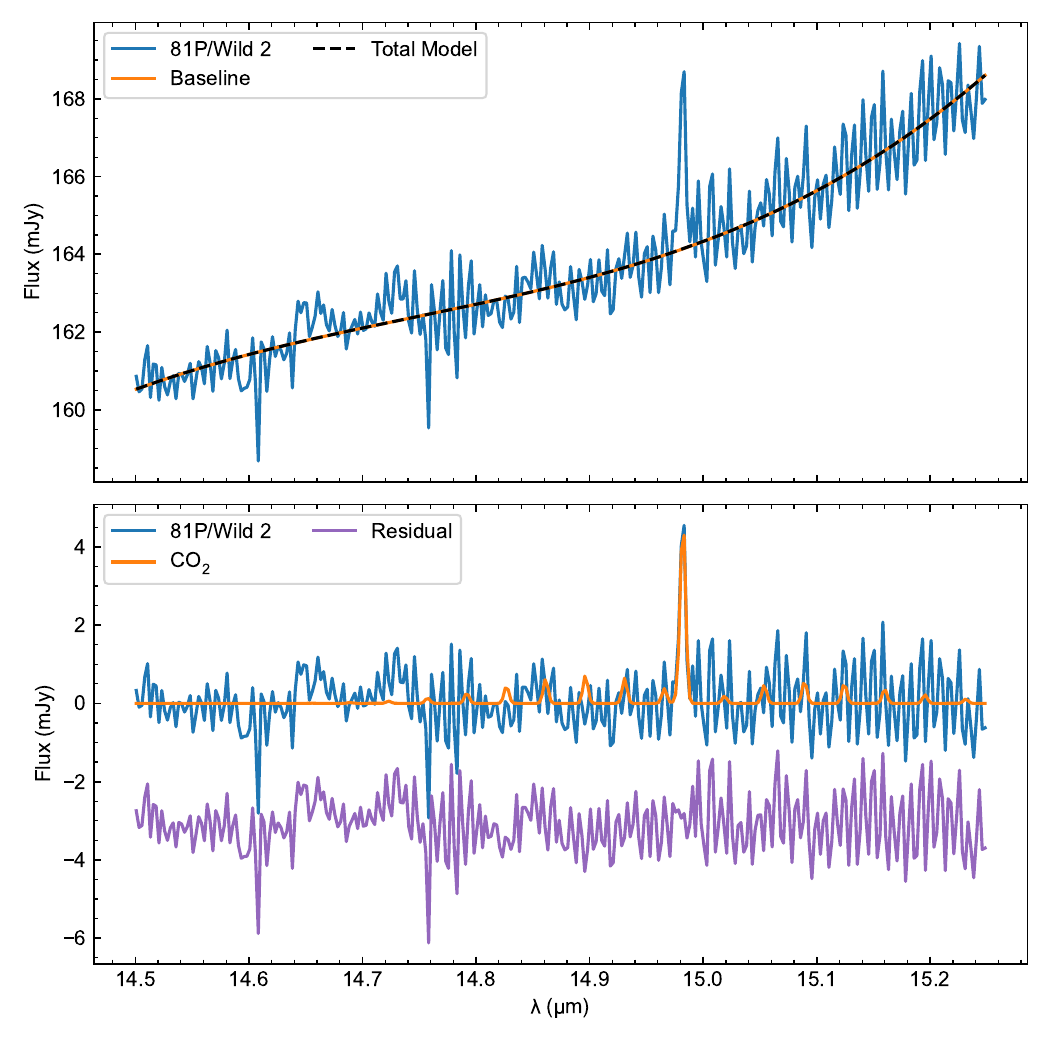}{0.45\textwidth}{(D)}
}
\caption{\textbf{(A)--(C).} As in Figure~\ref{fig:spec-panels1} for \ce{H2O}, OCS, CN, \ce{CH3OH}, \ce{H2CO}, \ce{C2H6}, \ce{CH4}, and OH* in 81P on March 24. \textbf{D.} As in Figure~\ref{fig:spec-panels1} for \ce{CO2} on March 20.
\label{fig:spec-panels2}}
\end{figure*}

Figures~\ref{fig:spec-panels1} and \ref{fig:spec-panels2} show extracted spectra and best-fit models for each detected species, and Table~\ref{tab:qs} gives the respective values of $Q$ and \trot{}. Table~\ref{tab:qs} shows that, compared to measured Jupiter-family comets, 81P was depleted in CO, \ce{CH4}, and \ce{CH3OH}, enriched in \ce{CO2}, and consistent with average values for all other species. Our simultaneous measure of \ce{^12CO2} and \ce{^13CO2} enabled us to derive a ratio $\ce{^12C/^13C}=85\pm4$, consistent with the terrestrial value of 89. The significant decrease (by near a factor of 3) of the \ce{CO2}/\ce{H2O} abundance from March 20 to March 24 is remarkable. Examination of Table~\ref{tab:qs} shows that $Q(\ce{H2O})$ rose by 38\% from one epoch to the next, while $Q(\ce{CO2})$ fell by 57\%, contributing to the dramatically enriched \ce{CO2} abundance on March 20 compared to more nearly median values on March 24. Although previous studies have found similar variations in molecular abundances measured at infrared wavelengths across apparitions \citep[e.g., 2P/Encke;][]{Roth2018}, such a dramatic decrease within the space of a few days is unusual. It is interesting to note that significant differences in \ce{CO2}/\ce{H2O} were found for interstellar comet 3I/ATLAS measured with JWST, depending on whether the NIRSpec \citep{Cordiner2026,Roth2026a} or MIRI \citep[measured several days after NIRSpec;][]{Belyakov2026} \ce{CO2} bands were analyzed. Future work with more nearly simultaneous measures of \ce{CO2} using MIRI and NIRSpec may shed additional light on comparisons between the 4.25 $\mu$m and 15 $\mu$m \ce{CO2} bands.

\cite{DelloRusso2014} characterized the volatile composition of 81P during its 2010 perihelion passage using high-resolution ground-based near-infrared spectroscopy. For molecules common to both studies, our molecular abundance for \ce{C2H6} is higher, whereas our abundances for HCN and \ce{H2CO} are lower, yet our abundance for \ce{CH3OH} is in formal agreement. However, all abundances between the two studies agree within $2\sigma$ uncertainty, indicating that there were no significant changes in the molecular abundances of these molecules over the course of three perihelion passages (in 2010, 2016, and 2022).

\begin{deluxetable*}{cccccc}
\tablenum{2}
\tablecaption{\label{tab:qs} Molecular Production Rates in 81P/Wild 2 and Comets Measured}
\tablewidth{0pt}
\tablehead{
\colhead{Molecule} & \colhead{$Q$\sups{(a)}}  & \colhead{\trot{}\sups{(b)}} & \colhead{$Q_x/Q(\ce{H2O})$\sups{(c)}} & \colhead{$\langle Q_x/Q(\ce{H2O})\rangle$\sups{(d)}} & \colhead{$Q_x/Q(\ce{H2O})$\sups{(e)}} \\
\colhead{} & \colhead{($10^{25}$ \ps{})} & \colhead{(K)} & \colhead{(\%)} &  \colhead{(\%)} & \colhead{(2010, \%)}
}
\startdata
\multicolumn{6}{c}{\textbf{2023 March 20, MIRI/MRS, \rh{}=1.87 au, $\Delta$=1.43 au}} \\
\ce{H2O} & $180\pm4$ & $24.0\pm0.4$ & 100 & ... & ... \\
\ce{CO2} & $80.8\pm10.5$ & $32\pm12$ & $45\pm6$ & $12\pm2$ & ... \\
\hline
\multicolumn{6}{c}{\textbf{2023 March 24, G395H/F290LP, \rh{}=1.85 au, $\Delta$=1.42 au}} \\
\ce{H2O} (2.9 $\mu$m) & $264\pm7$ & $38\pm3$ & 100 & ... & ... \\
\ce{H2O} (4.8 $\mu$m) & $272\pm8$ & (38) & $103\pm4$ & ... & ... \\
CO & $1.90\pm0.24$ & (38) & $0.72\pm0.09$ & $1.3\pm0.6$ & ... \\
\ce{CO2} & $44.8\pm1.3$ & $41.2\pm0.5$ & $17.0\pm0.2$ & $12\pm2$ & ... \\
\ce{^13CO2} & $0.53\pm0.02$ & $44\pm2$ & $0.20\pm0.01$ & ... & ... \\
\ce{CH4} & $1.20\pm0.06$ & $39\pm2$ & $0.46\pm0.02$ & $0.7\pm0.2$ & ... \\
\ce{C2H6} & $1.54\pm0.08$ & (38) & $0.59\pm0.03$ & $0.46\pm0.12$ & $0.45\pm0.05$ \\
\ce{CH3OH} & $2.88\pm0.24$ & $39\pm1$ & $1.09\pm0.08$ & $1.73\pm0.33$ & $0.9\pm0.3$ \\
\ce{H2CO} & $<0.2$ $(3\sigma)$ & (39) & $<0.09$ $(3\sigma)$ & $0.26\pm0.10$ & $0.22\pm0.08$ \\
HCN & $0.47\pm0.09$ & (38) & $0.18\pm0.04$ & $0.17\pm0.03$ & $0.27\pm0.03$ \\
OCS & $0.29\pm0.03$ & (38) & $0.10\pm0.01$ & $0.126\pm0.034$ & ... \\
CN & $0.25\pm0.03$ & (38) & $0.09\pm0.01$ & ... & ... \\
\enddata
\tablecomments{\sups{(a)}$Q$ is the molecular production rate. \sups{(b)} \trot{} is the rotational temperature. \sups{(c)} $Q_x/Q(\ce{H2O})$ are molecular abundances with respect to \ce{H2O} (this work). \sups{(d)}$\langle Q_x/Q(\ce{H2O})\rangle$ are average molecular abundances with respect to water in comets reported by \cite{Biver2024b,Harrington2022,Saki2020,Roth2020,Roth2017,DiSanti2017,DelloRusso2016}. \sups{(e)} Molecular abundances with respect to \ce{H2O} reported for 81P during its 2010 perihelion passage \citep{DelloRusso2014}.}
\end{deluxetable*}

\begin{figure}
\plotone{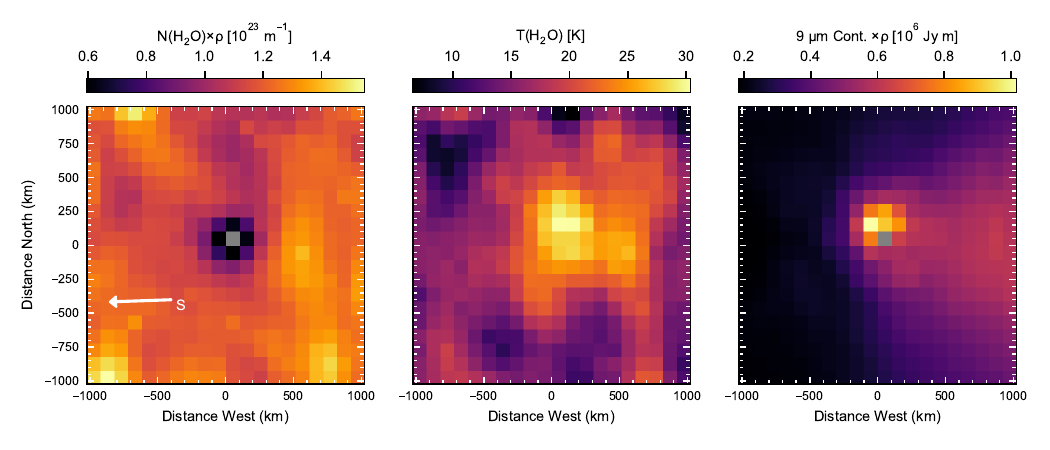}
\caption{\textbf{Left.} Map of $N\times\rho$ for \ce{H2O} in 81P on March 20 with MRS. The white arrow shows the projected direction of the Sun. \textbf{Middle.} Map of \trot{} for \ce{H2O}. A $3\times3$ boxcar smoothing kernel was applied during the retrieval to improve S/N. \textbf{Right.} Map of 9 $\mu$m projected continuum intensity ($I\times\rho$).
\label{fig:maps1}}
\end{figure}

\begin{figure}
\plotone{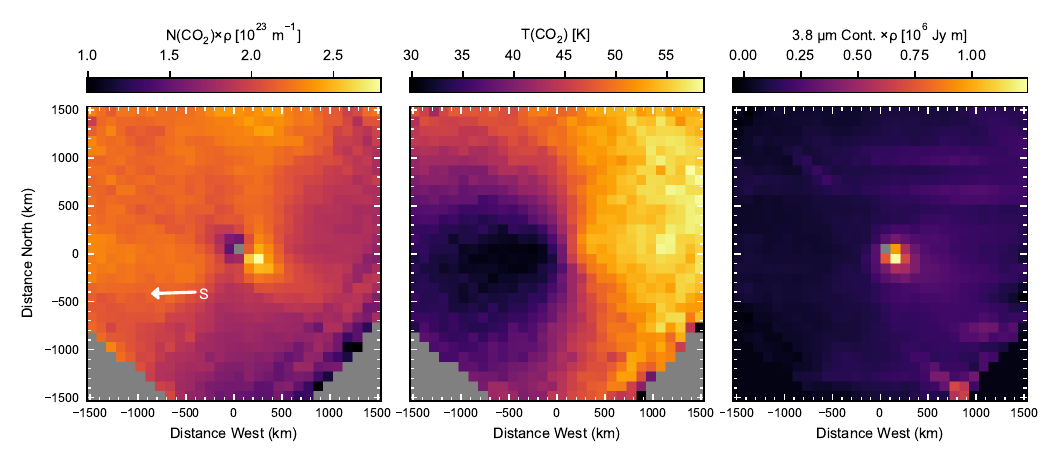}
\caption{\textbf{Left.} Map of $N\times\rho$ for \ce{CO2} in 81P on March 24 with NIRSpec. The white arrow shows the projected direction of the Sun. \textbf{Middle.} Map of \trot{} for \ce{CO2}. A $3\times3$ boxcar smoothing kernel was applied during the retrieval to improve S/N. \textbf{Right.} Map of 3.8 $\mu$m projected continuum intensity ($I\times\rho$).
\label{fig:maps2}}
\end{figure}

Our spaxel-by-spaxel maps of the brightest molecules, \ce{H2O} and \ce{CO2}, show considerable differences in the spatial distributions of column density ($N$) and \trot{} of each (Figures~\ref{fig:maps1} and \ref{fig:maps2}). To correct for expansion dilution effects, $N\times\rho$ is plotted. In such a plot, an isotropic outflow would appear approximately flat as a function of nucleocentric distance. Although at lower S/N, the MIRI \ce{H2O} $N\times\rho$ distribution is relatively flat with nucleocentric distance, indicative of an isotropic distribution. Likewise, \trot{} for \ce{H2O} peaks near the nucleus position and quickly falls off, although it is somewhat warmer in the anti-sunward direction. The simultaneously measured continuum emission shows a broad anti-sunward plume of dust. In contrast, the \ce{CO2} emission measured with NIRSpec is asymmetric, with a broad, approximately sunward fan of emission and a narrow anti-sunward jet. The rotational temperature shows a stark sunward/anti-sunward dichotomy, with the anti-sunward direction distinctly warmer. Furthermore, the overall \trot{} for \ce{CO2} is nearly twice as high than that for \ce{H2O}. The continuum in both the MIRI and NIRSpec observations shows an anti-sunward enhancement, although it is not as pronounced in the latter. 

These differences in column density and \trot{} distributions for \ce{H2O} and \ce{CO2} in 81P are similar to those measured by JWST for comet C/2017 K2 (PanSTARRS) and the interstellar object 3I/ATLAS \citep{2025PSJ.....6..139W,Roth2026a}. The lower \ce{H2O} temperatures were attributed to its more efficient rotational cooling compared to \ce{CO2} for both of these objects. Similarly, the lower temperatures for both molecules in the sunward direction were associated with more efficient adiabatic cooling in the sunward hemisphere of the coma. In addition to C/2017 K2 (PanSTARRS) and 3I/ATLAS, these sunward/anti-sunward dichotomies in \trot{} were also measured for \ce{H2O} in C/2022 E3 (ZTF) by JWST \citep{Foster2026} and also for \ce{CH3OH} in 46P/Wirtanen with ALMA and IRAM \citep{Biver2021,Cordiner2023}.

%% file: results-thermal.tex
\section{Coma Flux Derivation and Dust Thermal Modeling}\label{subsec:cfx-dtm}

\subsection{MRS Coma Flux Derivation} \label{subsubsec:dw-dust-mrs}

The analysis of the coma flux of comet 81P was conducted using spectral extractions derived from the 
pipeline reprocessed MRS IFU cubes (Section~\ref{sec:obs}). Three apertures, centered on the 
MRS cube photocenter (i.e., on the nucleus), were employed with aperture diameters of 1\farcs00, 
1\farcs22, and 1\farcs41, with equivalent areal coverages of $0.25\pi$, $0.372\pi$, and $0.50\pi$ square-arcseconds,
respectively, to derive the aperture-integrated observed flux. The Aperture Correction Factor (ACF) was 
applied to the NEATM model (Section~\ref{subsec:nucleus}) to yield the nucleus predicted aperture flux, 
accounting for PSF losses, for each of the three apertures across the four MRS grating settings.

The coma fluxes in each of the three apertures were then taken as the observed fluxes minus the nucleus 
predicted aperture fluxes. Between these three apertures, the coma fluxes in the 1\farcs22 and 1\farcs41
apertures were consistent in shape after scaling the 1\farcs22 aperture flux by 1.19 (for an ideally 
expanding coma, this ratio would equal the ratio of the aperture radii, $0.705/0.610 = 1.156$). The 1\farcs00 
diameter aperture was considerably less consistent in spectral shape compared to the 1\farcs22 
and 1\farcs41 diameter apertures, especially at wavelengths $\ge 18$~\micron. The nucleus-predicted 
aperture flux uncertainty was assessed by scaling the 1\farcs22 aperture flux to the 1\farcs41
aperture (multiplying by a factor of 1.19) and then taking the absolute value of its difference with the 1\farcs41 aperture spectrum. 
The coma flux error at each wavelength was then determined by root-mean-square of the instrumental 
error and nucleus predicted aperture flux uncertainty.

\begin{figure}[h]
\begin{center}
\includegraphics[trim=0.20cm 0.05cm 0.07cm 0.07cm, clip, width=0.45\textwidth]{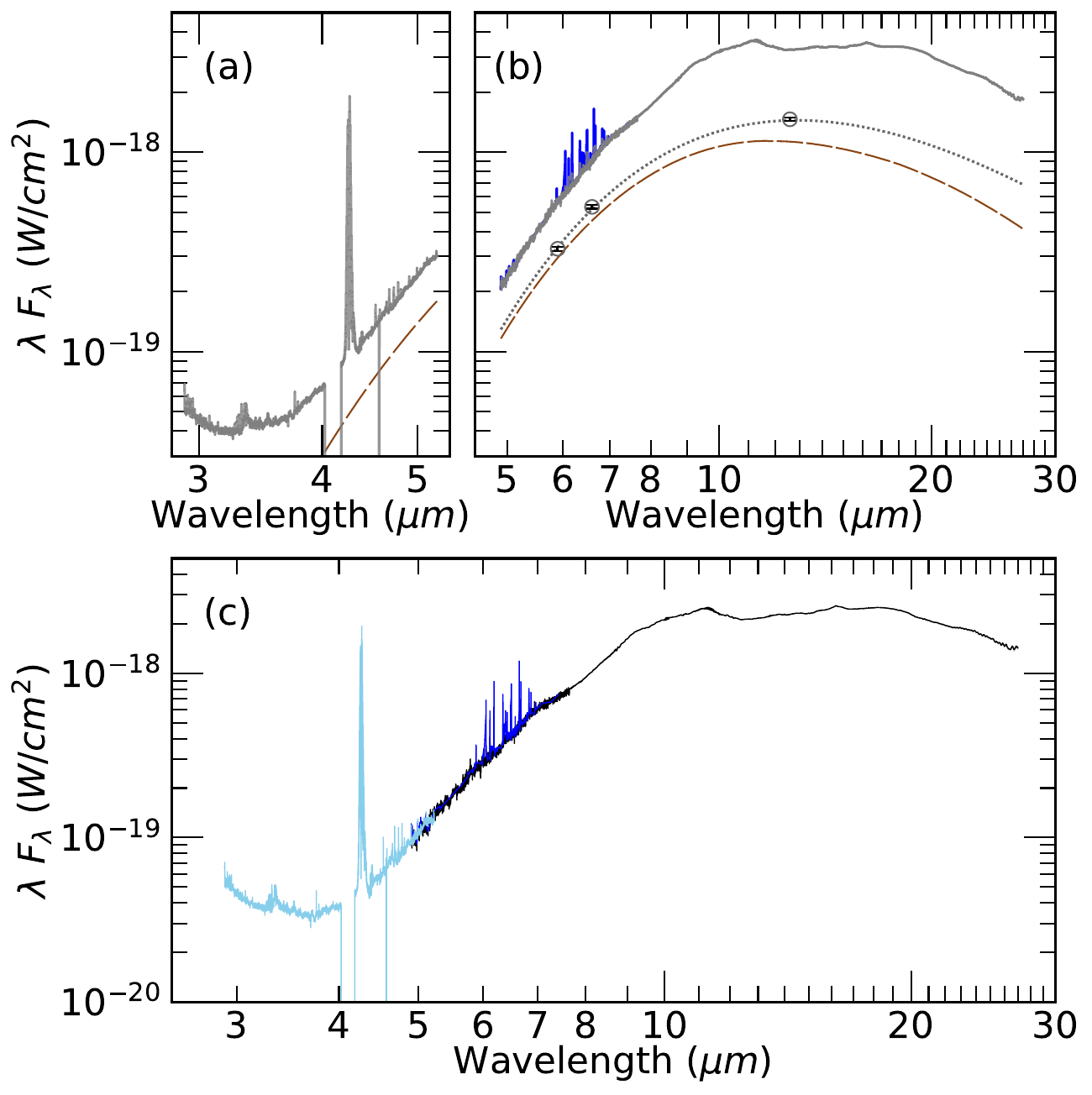}
\caption{Derivation of the $2.9$--$27$~\micron\ coma spectral energy
distribution ($\lambda F_\lambda$ vs.\ $\lambda$) for comet 81P/Wild~2
in a $1\farcs41$ diameter aperture. The measured fluxes and scaled
nucleus model fluxes ({\it saddlebrown} dashed lines) are shown in (a)
and (b), from which the nucleus-subtracted coma fluxes in (c) are
derived for NIRSpec ({\it skyblue}) and MRS ({\it black}). (a)~NIRSpec
flux ({\it dimgray}) with uncertainties ({\it silver}) plotted with
$0.96\times F_\lambda$(NEATM nucleus model) ({\it saddlebrown} dashed
line), where the $0.96$ scaling factor is set such that $1.042\times$
the NIRSpec coma flux matches the MRS coma flux in the overlapping
wavelength region (Section~\ref{subsubsec:dw-dust-nirspec}). 
(b)~MRS flux ({\it dimgray}) with uncertainties ({\it silver}), the NEATM 
nucleus model ({\it dimgray} densely dotted line) with the flux points used 
to derive the model (open circles), and the aperture-corrected nucleus model
(ACF$\times$NEATM, {\it saddlebrown} dashed line) that is subtracted
from the measured flux to derive the MRS coma flux. The prominent
H$_2$O emission lines at MRS wavelengths are shown ({\it blue}).
(c)~The nucleus-subtracted coma SED combining NIRSpec ({\it skyblue})
and MRS ({\it black}), used as input to the thermal model and scattered
light analysis (Section~\ref{subsubsection:dust-model}).  
\label{fig:PSJ-dwdust-1-20260528}}
\end{center}
\end{figure}

The aperture flux predicted for the nucleus for all three apertures presented a discontinuity 
(drop) at 17.979~\micron, which was most severe for the smallest aperture. This 
discontinuity was not seen in the data before nucleus 
subtraction. After subtracting the aperture flux predicted for the nucleus for the 1\farcs41 diameter aperture, 
the coma flux was multiplied by 1.042 at $\lambda \geq 17.979$~\micron. In the analysis that follows,
we adopt the 1\farcs41 diameter aperture nucleus-subtracted flux as the coma flux, with the propagated 
uncertainties. The aperture-integrated flux spectra and nucleus models from which
the coma flux is derived are shown in Figure~\ref{fig:PSJ-dwdust-1-20260528}.
In this aperture, the nucleus flux was approximately 0.75 of the total, so subtraction of the 
nucleus was critical to deriving the coma dust thermal emission spectrum. This coma flux was used for 
modeling and analyses of the coma dust and organics. 

\subsection{NIRSpec Coma Flux Derivation} \label{subsubsec:dw-dust-nirspec}

\begin{figure*}
\includegraphics[width=\textwidth]{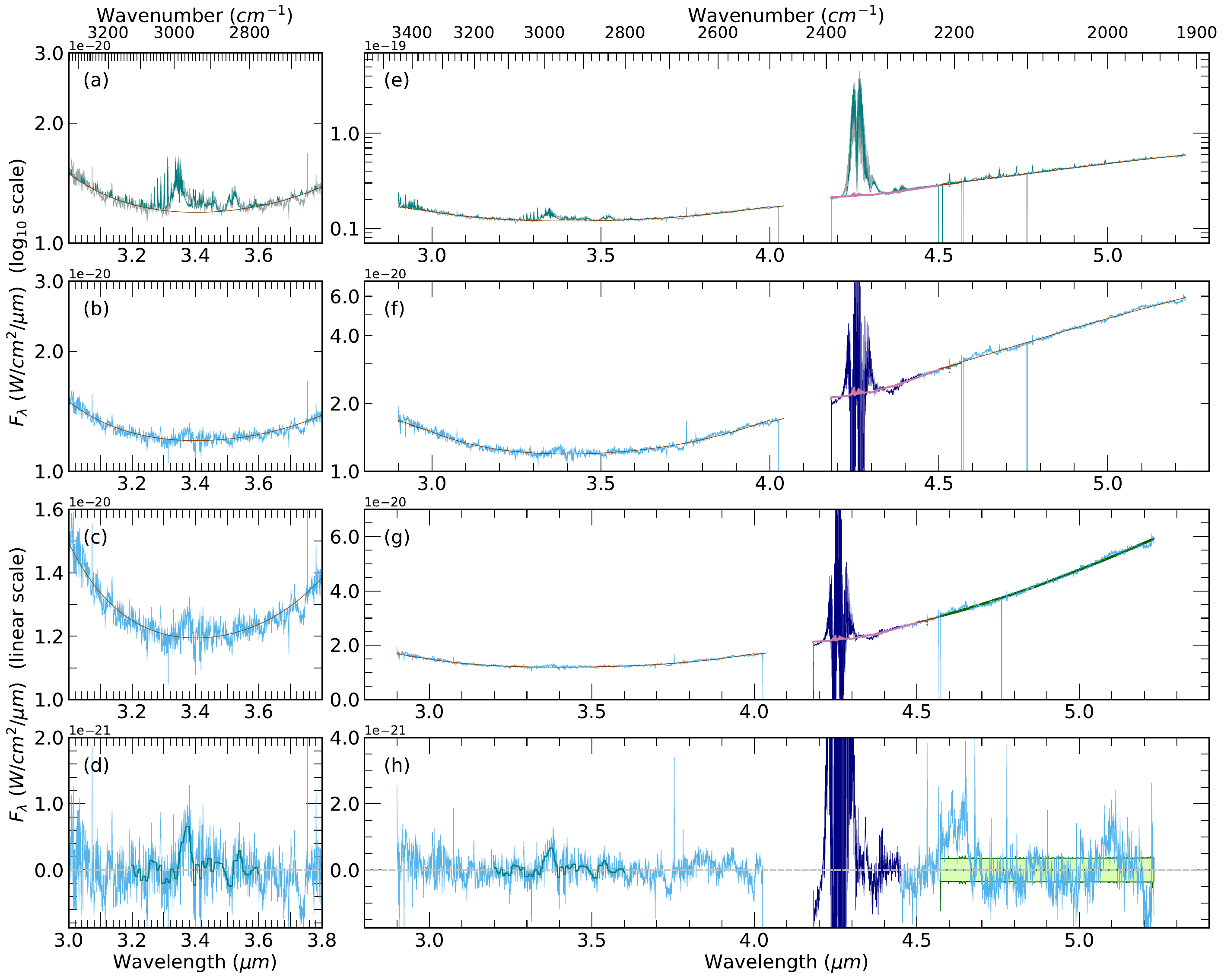}
\caption{Comet 81P/Wild~2's NIRSpec PSG model for gaseous molecular
emission lines and piecewise continuous baseline, and residual with
$3.42$~\micron\ feature, in a $1\farcs41$ diameter aperture. Left
panels (a--d) focus on the 
$3.0$--$3.8$~\micron\ 
region and right panels
(e--h) show the $2.9$--$5.23$~\micron\ wavelength range.
(a,~e) Measured flux density (coma plus nucleus) ({\it darkslategray})
plotted against the PSG model (PSG gas and PSG baseline) ({\it teal})
and the PSG baseline ({\it saddlebrown} and {\it palevioletred} for
the CO$_2$ region), on a $\log_{10}$ $y$-axis scale.
(b,~f) Flux density minus the PSG gas model ({\it skyblue} and
{\it navy} in the CO$_2$ region) compared to the PSG baseline
({\it saddlebrown} and {\it palevioletred} for the CO$_2$ region),
on a $\log_{10}$ $y$-axis scale.
(c,~g) The same data as (b,~f) on a linear $y$-axis scale, showing
the choice of the $4.57$--$5.23$~\micron\ continuum used in the
thermal modeling ({\it darkgreen} points and caps with {\it chartreuse}
error bars), which is the baseline coupled with an uncertainty envelope
encompassing a range of flux values and instrumental uncertainties in
this region where the PSG model lacks identified molecular species.
(d,~h) The PSG residual (measured flux minus the sum of the PSG
molecular line model and the PSG model baseline), on a linear $y$-axis
scale. The residual spanning $3.2$--$3.6$~\micron\ and centered at
$4.2$~\micron\ is median filtered ({\it teal}). The
$\lambda \geq 4.57$~\micron\ continuum uncertainties are more clearly
displayed in (h) compared to (g). 
The PSG gas and baseline were fitted on 2025 Aug 16, with the
$3.0$--$4.0$~\micron\ region refitted on 2026 Jun 11 (see text).
\label{fig:PSJ-dwdust-3}}
\end{figure*}

Spectral extractions in a 1\farcs41 diameter aperture were performed on the NIRSpec IFU spectral cubes. PSG models were run to fit the molecular line emission and continuum simultaneously. The molecular lines were subtracted to yield a residual containing scattered light, thermal dust emission, and features not fitted by the PSG model as described in Sections~\ref{subsec:h2o} and \ref{subsec:nirspec}. The PSG molecular emission model used for dust scattered light and thermal modeling was run on 2025 Aug 16. The CH$_3$OH model within the PSG, particularly the $\nu_9$ band, has since evolved \citep[see ][for further details on the challenges associated with modeling this band]{Villanueva2012a,Villanueva2025}. A revised PSG run on 2026 Jun 11 updated the \ce{CH3OH} model between $3.0$--$4.0$~\micron, while the baseline
changed negligibly ($\pm 2\%$) in that region (Fig.~\ref{fig:PSJ-dwdust-3}). Because the scattered light and thermal modeling used the gas-subtracted data from the 2025 Aug 16 PSG run, the PSG model outside $3.0$--$4.0$~\micron\ retains the original 2025 Aug 16 output.
The PSG molecular line model, the piecewise PSG baseline, and the
resulting residual are shown in Figure~\ref{fig:PSJ-dwdust-3}.

The nucleus NEATM model was extrapolated to NIRSpec wavelengths. Because of NIRSpec's higher 
spatial resolution, no ACF was applied, and the NEATM model directly yielded the nucleus total model 
flux. A simple subtraction of this flux was not feasible because the NIRSpec data were taken at a slightly 
different epoch than the MRS data. In the wavelength region of overlap with the MRS, the NIRSpec 
total flux (coma + nucleus) matched the MRS coma flux alone, indicating that the NIRSpec epoch 
measured slightly less coma flux than the MRS epoch.

Under the supposition that the same active areas have the same level of activity and contribute to the 
flux in the beam, an epoch change relates to a change in the projected cross-sectional area of the 
nucleus, with a similarly proportional change in the coma contribution. Therefore, we scaled the nucleus 
NEATM model by 0.96$\pm$0.01, which is commensurate with a small change in the projected area 
of the nucleus size due to the comet's rotation. We then subtracted the scaled nucleus flux to 
derive the coma flux, and then multiplied the coma flux by the inverse factor 1.042$\pm$0.011.
This factor was chosen such that after subtraction, the NIRSpec coma flux matched the MRS coma 
flux in the overlap region. 

\subsection{Iterative Modeling of the Dust Thermal Emission and Scattered Light} \label{subsubsection:dust-model}

The dust thermal model and the scattered light model were fitted iteratively in flux density units,
 $F_{\lambda}$ (W~cm$^{-2}$~$\mu \text{m}^{-1}$). Throughout this section, ``flux'' means $F_{\lambda}$.  The H$_{2}$O molecular 
emission-line model (Section~\ref{subsec:h2o}) was subtracted from the MRS Channel~1 data at the H$_{2}$O line 
positions. However, the subtraction was imperfect and left residuals. To remove these residuals, the 
line-to-continuum ratio was assessed at each H$_{2}$O line position, points with line-to-continuum $\ge$3 
were masked, and a median filter was applied to derive a new continuum. This mask-and-filter procedure 
was repeated three times (see Appendix~\ref{appendix1}). The resulting continuum, free of narrow-line residuals, 
was used in preparing the residuals shown in Figure~\ref{fig:PSJ-dwdust-4-20260528}.

Thermal emission contributes significantly to the longer wavelengths of NIRSpec and scattered light 
contributes minimally to the shorter wavelengths of MRS ($\lambda \ltsimeq 8$~\micron). The region of 
overlap between MRS and NIRSpec is near where the thermal emission and scattered light have 
approximately equal contributions. The thermal emission flux falls sharply towards shorter 
MRS wavelengths ($\lambda \ltsimeq 8$~\micron) and the scattered light gradually decreases in flux towards longer NIRSpec
wavelengths ($\lambda \gtsimeq 4$~\micron). To achieve the best thermal model that would extend to 
NIRSpec wavelengths, the scattered light subtracted spectra were together fitted down to 4.1~\micron. Potential polycyclic aromatic hydrocarbon (PAH) emission in MRS starts at $\sim$5.5~\micron \  \citep[e.g,][]{2025PSJ.....6..139W}, 
yielding the 4.9--5.5~\micron{} region available in principle for establishing the sum of the scattered light 
and thermal emission for MRS. 


Comets contain refractory dust grains that are a mixture of ISM dust and solar nebula processed material. 
The grains are composed primarily of silicates (amorphous  and crystalline), that produce distinct emission 
features in the mid-IR, and dark  absorbing material that produces a featureless underlying continuum.  
Hence, a dust thermal model is fitted to the coma flux minus the flux of scattered light from the coma. The dust thermal 
model \citep{2023PSJ.....4..242H, 2024come.book..577E, 2025PSJ.....6..139W} employs five 
materials: Mg:Fe = 50:50 amorphous silicates, Mg-rich crystalline silicates of olivine and orthopyroxene 
composition (Mg$_{2y}$Fe$_{2(1-y)}$SiO$_4$, $0.9 \leq y \leq 1$) and (Mg$_x$Fe$_{1-x}$SiO$_3$, $x \simeq 1$), 
and amorphous carbon \citep[glassy carbon,][]{1983PhDT........18E} that is featureless and highly absorbing. 

Our dust thermal model \citep[][and references therein]{2023PSJ.....4..242H} predicts the thermal emission per 
particle for an ensemble of particles with discrete compositions characterized by a differential size distribution, 
specifically the Hanner grain size distribution \citep[HGSD,][]{1994ApJ...425..274H}, defined by a slope $N$, 
the particle size at which the size distribution peaks ($a_p$), and a parameter $D$ for the porosity prescription 
$P = 1-f$, where $f$ is the filling factor and $f = (a/a_0)^D$ with $a_0 = 0.1$~\micron. The dust thermal 
model parameters that result from fitting the thermal model to the data by minimizing $\chi^2$ are the three 
HGSD parameters and the relative mass fractions of the five dust materials.

We used $\chi^2$ to assess the dust thermal model parameters for a particular model. We used 
the Akaike Information Criterion \citep[AIC,][]{doi:10.1177/0049124104268644}  to compare the 
dust thermal model fitted to different renditions of the data. We changed the weights of the coma 
flux uncertainties (from $1\times$ to $40\times$) in some spectral regions of MRS to improve the 
AIC. Specifically, including the 12.5--16.5~\micron{} wavelength
range with the instrumental errors always increased the AIC (worsened the fit). Increased uncertainties were 
applied to MRS between 12.5 and 16.5~\micron{} because of an unknown
dust component emitting prominently in this wavelength range. Increased uncertainties were also applied
in the short-wavelength shoulder of the silicate feature because of the potential presence of PAH emissions. Flux in the 5.60--8.65~\micron{} region was used to constrain the thermal model. However, this region has 
potential PAH emissions that are weak in contrast to the thermal emission. The MRS data in Channel~1 
has significant fringing or undulations in the pipeline produced data products \citep[e.g.,][]{2026AJ....171..304L}. 
For the thermal modeling, this was not median filtered.

\begin{figure}[h]
\begin{center}
\includegraphics[trim=0.20cm 0.05cm 0.07cm 0.07cm, clip, width=0.45\textwidth]{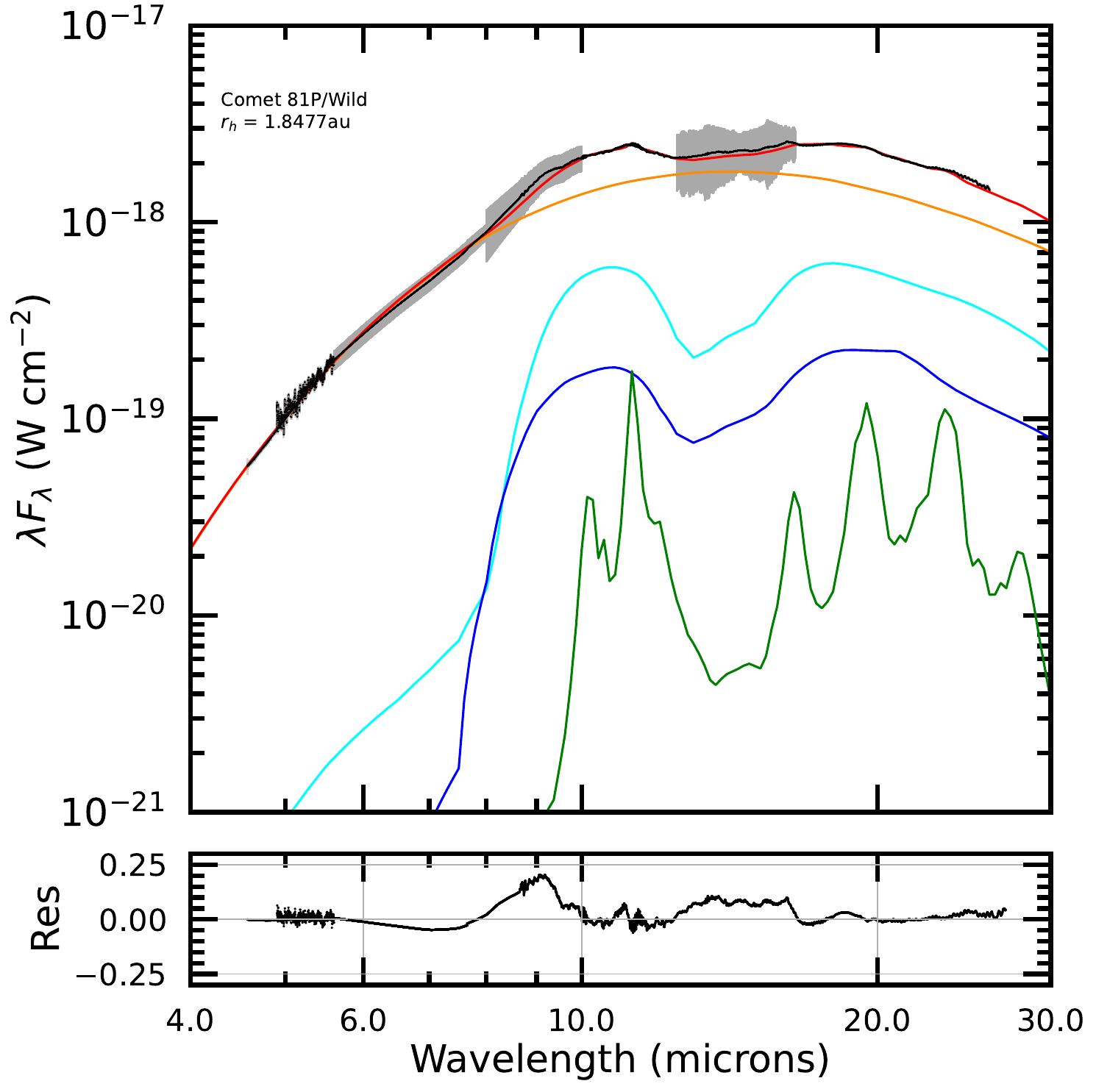}
\caption{Best-fit thermal model for the scattered light-subtracted
coma of comet 81P/Wild~2 in a $1\farcs41$ diameter aperture. The
flux points and uncertainties used to fit the thermal model are shown
({\it black}, {\it dimgray}) along with the residual. The thermal
model total ({\it red}) is the sum of dust components of amorphous
carbon (AC, {\it darkorange}), amorphous olivine (AO50, {\it cyan}),
amorphous pyroxene (AP50, {\it blue}), and crystalline olivine
(CO, {\it forestgreen}). A pie diagram of the mass fractions of these
dust components is in Figure~\ref{fig:PSJ-dwdust-5-20260528}. Enhanced
uncertainties are applied in regions of the spectrum where organics or
unknown carriers contribute to the observed emissions. Specifically,
for $5.6$--$8.65$~\micron\ an estimated baseline and range for the
uncertainties in that baseline were substituted for the measured values
and used in the thermal model fitting. This $5.6$--$8.65$~\micron\
estimated baseline undershoots the thermal model and the residual shows
an unusual shape in this region. The thermal model is adopted for the
$5.6$--$8.65$~\micron\ baseline and is also adopted for the baselines
in the $8.65$--$10$~\micron\ and $12.5$--$16.5$~\micron\ regions.}
\label{fig:PSJ-dwdust-2-20260558}
\end{center}
\end{figure}

A combination of two estimates of the local continuum was determined for under the 5.60--8.65~\micron{} 
PAH bands in order to do the thermal model fitting. Using an initial run for the thermal model, one 
continuum estimate was determined by shifting the amorphous carbon to lie under the 
PAH emission bands. A second continuum estimate was determined by shifting the amorphous 
olivine, which has a steeper spectral decline going towards shorter wavelengths in comparison 
to the amorphous carbon. A value of $2.5\times10^{-21}$~W~cm$^{-2}$~$\mu \text{m}^{-1}$ 
was 
assessed for the level of the pixel-to-pixel undulations in this region of the spectrum. This uncertainty 
was considered for each of the two local continua. The envelope of the upper and lower bounds of the 
combination of the two local continua was assessed and the mid-point determined. These mid-points 
and their uncertainties were used as constraints for the thermal model fitting in the 5.60--8.65~\micron{} 
wavelength range. This is shown in Figure~\ref{fig:PSJ-dwdust-2-20260558} for the last 
iteration of the thermal model.

\begin{figure}[h]
\begin{center}
\includegraphics[trim=0.07cm 0.05cm 0.07cm 0.07cm, clip, width=0.45\textwidth]{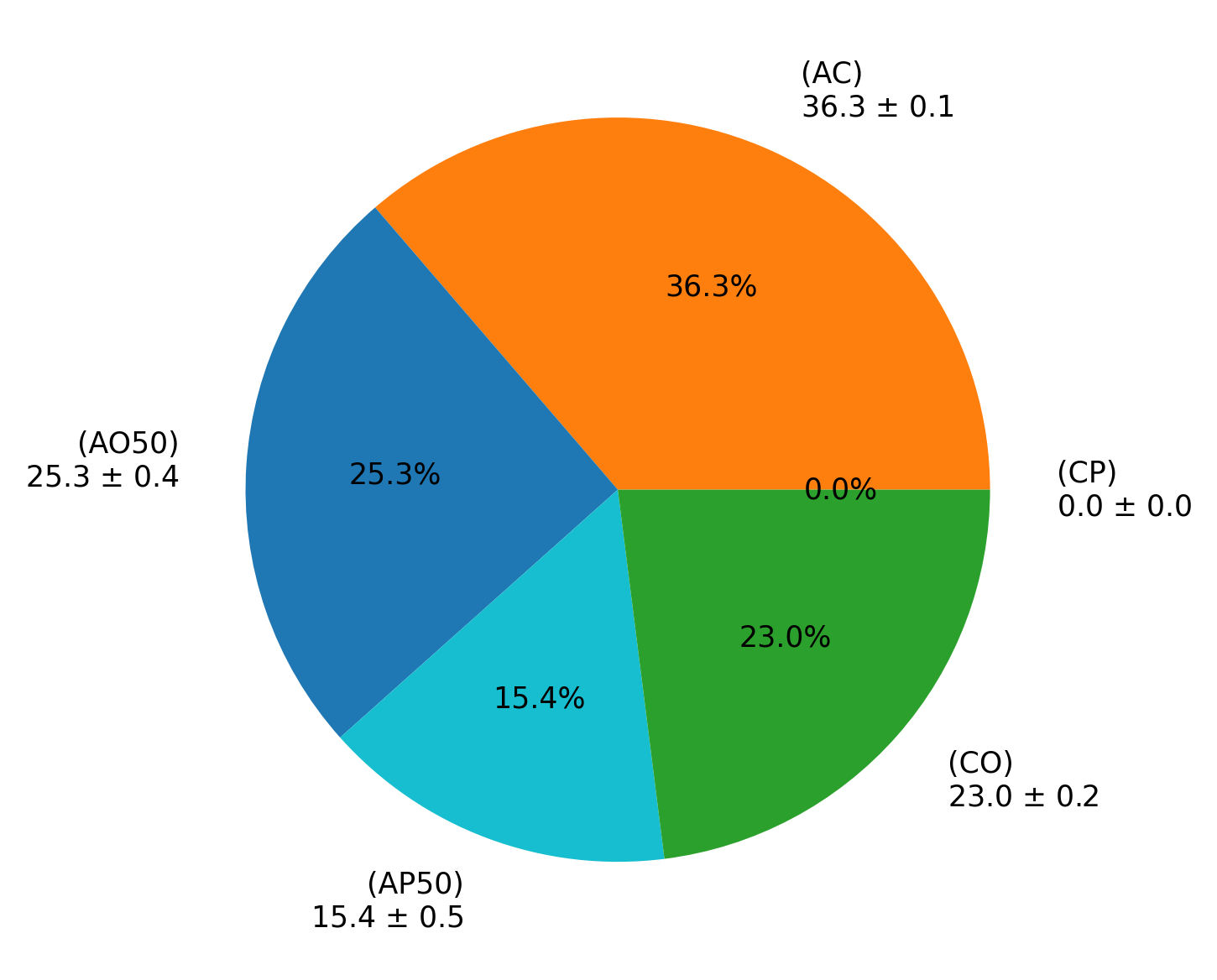}
\caption{The relative mass fraction for the submicron- to micron-size portion of the particle differential 
size distribution in the inner coma of comet 81P. The compositional designations illustrated in
the figure are: amorphous carbon (AC), amorphous silicates (AO50, AP50), and crystalline silicates (CO, CP).}
\label{fig:PSJ-dwdust-5-20260528}
\end{center}
\end{figure}

\begin{figure*}[h]
\begin{center}
\includegraphics[trim=0.20cm 0.05cm 0.07cm 0.07cm, clip, width=0.85\textwidth]{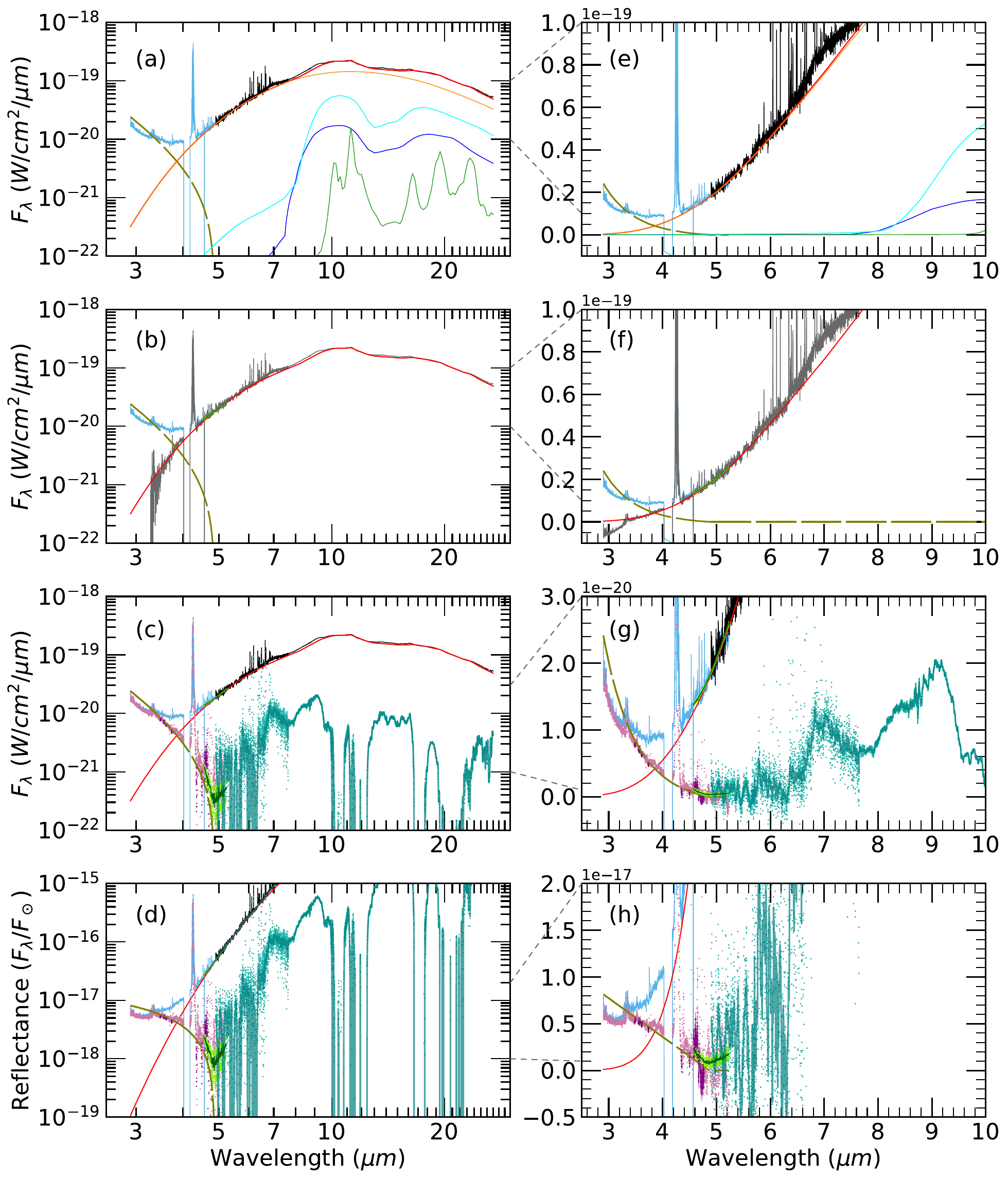}
\caption{Comet 81P/Wild~2's coma spectrum separated into components
derived from iteratively fitting the scattered light and the thermal
model. Panels (a--d) have 5-decade logarithmic $y$-scales and (e--h)
are zoomed-in linear inset plots. (a) Coma flux of NIRSpec
({\it skyblue}) and MRS ({\it black}) plotted with the scattered light
model ({\it olive} long dashed line) along with the thermal model total
({\it red}) and its dust components of amorphous carbon
(AC, {\it darkorange}), amorphous olivine (AO50, {\it cyan}), amorphous
pyroxene (AP50, {\it blue}), and crystalline olivine
(CO, {\it forestgreen}). (b) Coma flux minus the scattered light
({\it dimgray}) compared to the thermal model total ({\it red}).
(c) The thermal model residual (coma minus thermal model). The thermal
model residual for MRS is shown by data points ({\it lightseagreen})
and as a median filtered spectrum ({\it teal}). The scattered light
model ({\it olive} long dashed line) is compared to the thermal model
residual at NIRSpec wavelengths ({\it palevioletred}). The estimated
PSG baseline that was an input to the thermal model fitting is also
shown ({\it darkgreen} points and {\it chartreuse} error bars; see
Figure~\ref{fig:PSJ-dwdust-3}(g,~h)). (d) The scattered light model
({\it olive}) compared to the thermal model residual in Reflectance
units. The {\it purple} points highlight the NIRSpec data points used
in the scattered light model fit. (h) The scattered light linear model fit (
Reflectance), which passes through the lower error bars of the estimated baseline's
uncertainties ({\it chartreuse}) that reference the PSG model baseline
(Figure~\ref{fig:PSJ-dwdust-3}(h)).}
\label{fig:PSJ-dwdust-4-20260528}
\end{center}
\end{figure*}

The thermal model was fit to data that included MRS wavelengths from 5.60--8.65~\micron{} as well as NIRSpec 
wavelengths, with NIRSpec providing a fit to the coma continuum after subtracting an estimate of the 
scattered light. The thermal 
model had to span NIRSpec wavelengths because of the strong dust contribution to the coma aperture flux. In the NIRSpec wavelengths, there is significant scattered light declining toward longer wavelengths. 
A linear model was adopted for the scattered light in normalized reflectance \citep[flux divided by a scaled solar spectrum 
using `E490\_2014LR' from the \texttt{sbpy.calib.Sun} package;][]{Mommert2019}, with the 3.6~\micron{} flux 
point chosen as the point of normalization 
(at the end of the 3.2--3.6~\micron{} wavelength region considered for potential PAH emission in NIRSpec). 
The slope and intercept were iteratively determined using the SciPy \citep{2020NatMe..17..261V}
package \texttt{scipy.optimize.curve\_fit} so that the intercept was unity at 3.6~\micron{} in normalized reflectance. 
The NIRSpec scattered light was estimated by fitting under the molecular bands --- including 
under CO$_{2}$ (to which PSG did not fit uniquely due to opacity effects, see Section~\ref{subsec:nirspec})
and under weak lines not modeled by PSG --- using the flux minima as the fit points (Figure~\ref{fig:PSJ-dwdust-3}(g, h)). 
The slope was constrained such that the NIRSpec coma flux minus the scattered light and the 
MRS coma flux minus the scattered light agreed within the uncertainties with the thermal model across 
these wavelengths. 

The fit to the NIRSpec scattered light used select coma minus thermal model points that appear to be the 
lowest flux points in the ranges 3.48--3.745~\micron\ (3.48--3.50, 3.55--3.58, 3.60--3.62, 
3.727--3.745~\micron) and 4.345--4.77~\micron\ (4.345--4.360, 4.495--4.505, 4.570--4.770~\micron), 
together with the 4.57--5.2~\micron\ PSG baseline with increased uncertainties 
(dark green points and chartreuse error bars; Fig.~\ref{fig:PSJ-dwdust-3}(h)). These 
increased uncertainties in the 4.57--5.2~\micron\ PSG baseline were assessed from wavelength 
points where the PSG molecular model flux divided by the instrumental error was $\leq 5$, i.e., where no 
strong lines were present in the PSG molecular model. The uncertainties adopted were the root-mean-square 
of the instrumental uncertainties and the standard deviation 
($\sigma$ = 3.226~$\times$~10$^{-22}$~W~cm$^{-2}$~\micron$^{-1}$) of a Gaussian fitted 
to the histogram of the residual minus the Savitzky--Golay-smoothed residual (\texttt{scipy.signal.savgol\_filter}) residual.
The NIRSpec scattered light was then extrapolated to MRS, subtracted, and a new thermal model 
was assessed (i.e., the iterative process). 

Thermal models also were run and scattered light slopes determined for the combined NIRSpec and MRS
data with two additional NIRSpec scaling factors, the upper and lower scaling factors were related to
$1.042 \pm 0.011$ (see Section~\ref{subsubsec:dw-dust-nirspec}). The scattered light slopes, intercepts, and 
$R_{3.6~\micron}$ changed to $-0.693 \pm 5.0 \times 10^{-6}$~\micron$^{-1}$, $1.0 \pm 1.4 \times 10^{-6}$, and
$5.35 \times 10^{-18}$, respectively, for the upper scaling factor (1.053) and to
$-0.888 \pm 3.4 \times 10^{-3}$~\micron$^{-1}$, $1.0 \pm 1.1 \times 10^{-3}$, and
$5.134 \times 10^{-18}$, respectively, for the lower scaling factor (1.031). Even though the scattered light
slopes changed for NIRSpec, given that the scattered light does not extend far into MRS, the
corresponding thermal models for these two alternate NIRSpec coma scaling factors and scattered
light models differed primarily in the 4.1--5.2~\micron\ region. For the NIRSpec coma scale factor
of 1.053, the shallower scattered light slope, when extrapolated to MRS wavelengths, yields a
slight over-subtraction at 4.5--5.0~\micron\ in the overlap region between the two instruments.
For 1.031, in the overlap region the scattered light fit runs through the lower error bars of the
4.57--5.2~\micron\ PSG baseline with enhanced uncertainties, similar to 1.042, but its steeper
scattered light slope yields a slight over-subtraction at $\lambda > 5.0$~\micron. The 1.042
scaler yields the best total residual (coma flux minus the sum of the thermal model and the
scattered light model) in the overlap region. The total residual falls within the lower uncertainty
bounds of the 4.57--5.2~\micron\ PSG baseline's expanded uncertainties, which represents the best
determination of the local continuum beneath potential emissions not identified by the PSG (Fig.~\ref{fig:PSJ-dwdust-3}(h,g)).
The uncertainties from the scattered light and thermal models are not formally propagated into the
errors. The thermal model is highly non-linear and formal error propagation is beyond the scope of
this work. Nevertheless, the differences among the total residuals for the three NIRSpec scaling
factors, which exceed the internal uncertainties in the thermal models themselves, remain small
compared to the MRS instrumental uncertainties. The final scattered light fit and the corresponding
thermal model for the adopted NIRSpec coma scaling factor of 1.042 were used for the 1\farcs41
diameter aperture. The thermal model and the scattered light model, derived iteratively, and their
subtraction from the coma flux are shown in Figure~\ref{fig:PSJ-dwdust-4-20260528}.

\input{P81-dust-model-table1.tex}
\input{P81-dust-model-table2.tex}

\section{Comet 81P Coma Dust Composition} \label{sec:dw-dust-bestfit}

The full thermal model fit is shown in Figure~\ref{fig:PSJ-dwdust-2-20260558}.
Table~\ref{tab:composition-modelparams} lists the best-fit model parameters in terms of the
particle HGSD differential size distribution. Table~\ref{tab:mod-massfractions} lists the relative mass fractions 
of the submicron particles in that distribution, and Figure~\ref{fig:PSJ-dwdust-5-20260528} displays these mass fractions as a 
pie diagram. Mg-rich crystalline orthopyroxene is included in the fitting but yields
$N_p$(CP)$=0.0000\pm0.0000$, and so is omitted from 
Tables~\ref{tab:composition-modelparams} and~\ref{tab:mod-massfractions}.

In the thermal model, the crystalline silicates are confined to the 0.1--1~\micron{} size range, and the mass 
fractions of the other compositions are tabulated over the same range. This range is chosen because 
crystals computed at radii larger than 1~\micron{} have resonances with long-wavelength 
shoulders and lower contrast that do not fit the observed resonances \citep{2013ApJ...766...54L, 2024come.book..577E}.
Porous amorphous particles out to 100~\micron{} radii are included in the HGSD. The best-fit fractal 
dimension $D=2.727$ is considered `moderately porous' and is characteristic of the thermal-model 
outcomes for amorphous silicates and amorphous carbon in many comets \citep{2023PSJ.....4..242H}. 
The peak grain size $a_{p}=0.7$~\micron{} lies in the mid-range of values reported for 22 comets 
in a Spitzer survey modeled over mid-IR through far-IR wavelengths  \citep[7.5--30~\micron;][Figure~5a]{2023PSJ.....4..242H}; 
however, the slope $N=4.3$ is significantly steeper. A steeper slope emphasizes the submicron particles 
relative to shallower slopes (e.g., $N=3.5$). 

Amorphous Mg:Fe olivine-type and Mg:Fe pyroxene-type silicates dominate the relative mass fractions 
of the submicron silicates. The Mg-rich crystalline olivine shows a weak resonance and a significant mass 
fraction. This weak resonance is commensurate with the lower temperatures per particle that arise from 
the lower absorptivity at high Mg-content. The null contribution from Mg-rich crystalline orthopyroxene 
noted above likely extends to Mg-rich crystalline clinopyroxene, since the two have similar spectral 
shapes at high Mg-content \citep{2002A&A...391..267C}.

\begin{figure}[ht!]
\begin{center}
\includegraphics[trim=0.07cm 0.05cm 0.07cm 0.07cm, clip, width=0.75\textwidth]{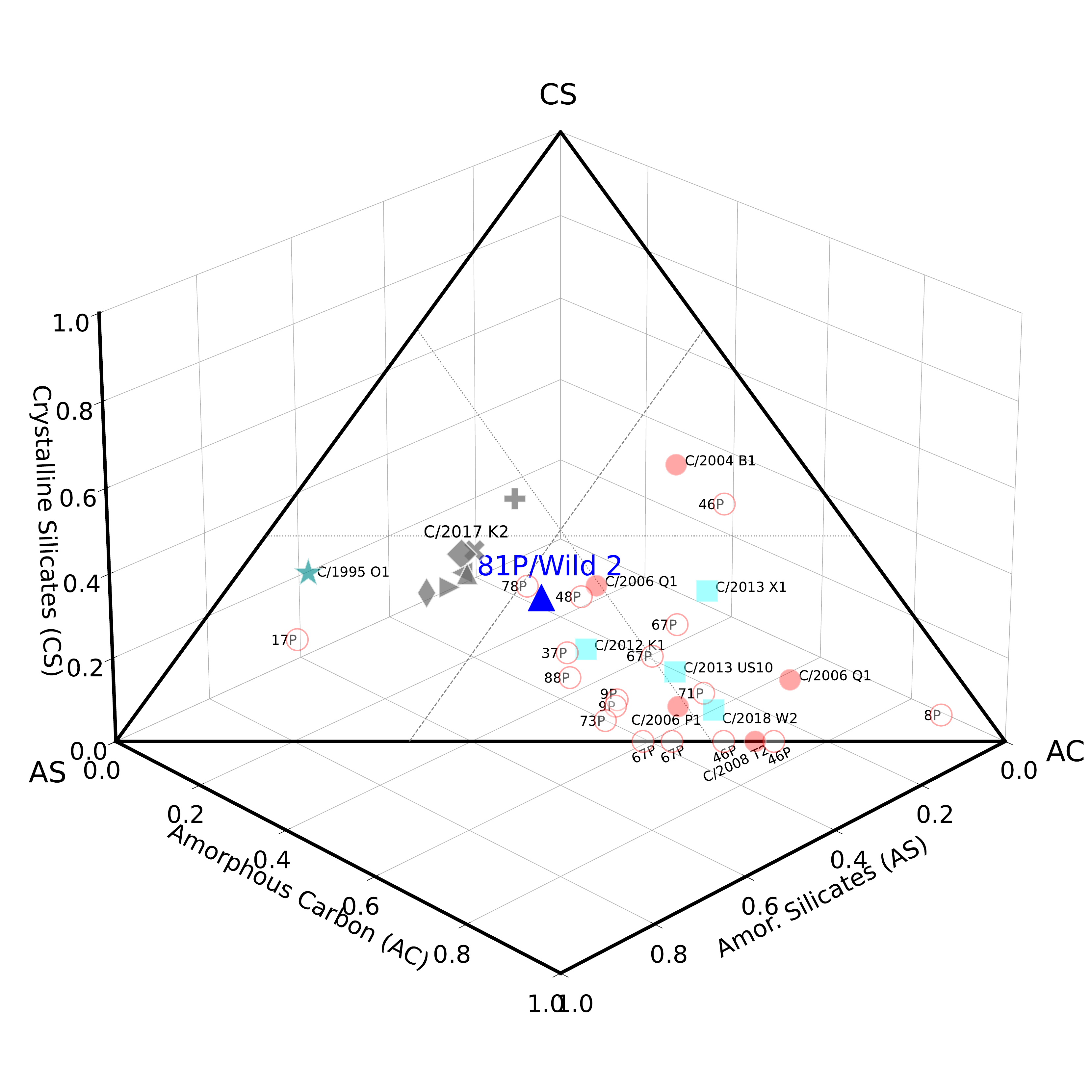}
\caption{A ternary diagram of the relative mass fractions of the
thermal dust model components for comet 81P/Wild~2 (large blue
triangle) and a comparison sample of comets from the Spitzer survey
\citep{2023PSJ.....4..242H} and comet C/2017~K2 (PanSTARRS) observed
by JWST \citep{2025PSJ.....6..139W}. Solid symbols are Oort cloud
comets; open symbols are Jupiter Family comets. The three axes
represent the summed mass fractions of amorphous silicates
(AS $=$ AO50 $+$ AP50), amorphous carbon (AC), and crystalline
silicates (CS $=$ CO $+$ CP), where AS $+$ AC $+$ CS $= 1$. A ternary
diagram represents compositions in which three components sum to unity
as a triangular plane in three-dimensional composition space; the
axial grid is included to highlight this three-dimensional geometry.}
\label{PSJ-dwdust-6-20260528.pdf}
\end{center}
\end{figure}

Figure~\ref{PSJ-dwdust-6-20260528.pdf} shows the composition of comet 81P alongside other comets in the 
coordinates of the mass fractions of amorphous carbon (AC), amorphous silicates (AS, the sum of amorphous 
olivine and pyroxene), and crystalline silicates (CS, the sum of crystalline olivine and orthopyroxene). 
The comparison sample includes the Spitzer survey comets modeled over the full wavelength 
range \citep[7.5--30~\micron;][]{2023PSJ.....4..242H} and comet C/2017 K2 (PanSTARRS), 
observed by JWST \citep{2025PSJ.....6..139W}. Comet 81P is more silicate-rich than most 
comets in this figure, and its crystalline mass fraction,
$f_{\rm{cryst}}=f(CO)/[f(CO)+f(AP50)+f(AO50)]$ (cf.\ Table~\ref{tab:mod-massfractions}), is as high as or 
higher than most other comets analyzed applying the same methodology. Only comet C/1995~O1~(Hale-Bopp), 
comet 17P/Holmes after outburst, comet 78P/Gehrels~2, and comet C/2017~K2~(PanSTARRS) are more 
silicate-rich. It so happens that like comet 81P, comet 78P/Gehrels~2 and C/2017~K2~(PanSTARRS) 
have 14.5~\micron{} features \citep{2025PSJ.....6..139W, 2023PSJ.....4..242H}. Hale-Bopp and 17P/Holmes had 
very prominent silicate resonances with distinct crystal features because their comae were silicate-rich 
and their particles were very small, with $a_p=0.2$~\micron{} and 0.3~\micron, 
respectively \citep{2002ApJ...580..579H, 2023PSJ.....4..242H}. Smaller particles have 
higher-contrast resonances. Comet 81P thus belongs to a small group of silicate-rich, crystal-rich 
comets that also includes Hale-Bopp, 17P/Holmes (after outburst), 78P/Gehrels~2, and  C/2017~K2~(PanSTARRS).

\vspace{0.2cm}
\section{Mineralogy: Stardust versus JWST} 
\label{sec:dw-dust-stardust}

\subsection{The Submicron Stardust Components}\label{sec-dw-dust-stardust-partA}

In this section we examine the major mineral characteristics of the Stardust samples in order 
to compare them with the JWST thermal model results for the dust in 
comet 81P/Wild 2's coma \citep{2014AREPS..42..179B, 2024come.book..577E}. The thermal 
model preferentially samples the smallest particles in the size distribution because they produce 
the higher-contrast resonances. The mineralogies of olivine crystals as large or larger than 
$\sim$2~\micron{} radii have resonances that are broadened towards
longer wavelengths and are filled in between the far-IR resonances \citep{2013ApJ...766...54L} to
the extent that their resonances disagree with observed IR emission spectral features \citep[e.g., by
comparison to the distinct resonances from Mg-rich olivine submicron crystals in comet C/1995 O1
(Hale-Bopp);][]{1997Sci...275.1904C}. The absorption spectrum of a 20~\micron{} 
anhydrous IDP also shows this trend where the mid-IR resonances become saturated
\citep[Figure~1,][]{2003LPI....34.1903K}.  The submicron-to-micron particle radii range is therefore where the JWST
thermal model and the Stardust laboratory characterization can be cross-compared. 
Thus, studies of fine-grained materials (FGM) in Stardust samples are the focus of this section. 

In Stardust samples, the submicron components of the captured particles were disaggregated into the
bulbs of some tracks during the 6~km~s$^{-1}$ flyby capture.  The micron-sized and larger components
penetrated deeper into the aerogel.  The carrot-shaped and rutabaga-shaped tracks had these large
terminal mineral particles. The terminal particles are well studied \citep{2024SSRv..220...79Z,
2024come.book..577E} and are dominated by Mg:Fe olivines and pyroxenes. The olivines have a
wide-and-flat or bimodal Mg distribution, spanning from Mg-rich LIME olivines
\citep{2017M&PS...52.1612J} to Mg$\sim$0.5 and one nearly pure fayalite grain was also found
\citep[Mg$\sim$0;][]{2024SSRv..220...79Z}. Most terminal olivines are 5--30~\micron{} in diameter, with
minor-element abundances commensurate with mostly type II (Mg$\leq$0.9) chondrules and some type I
chondrules (0.9$\leq$Mg$\leq$1.0). These chondrules are of igneous origin, formed by rapid heating
and melting \citep{2014GeCoA.142..240F, 2017RSPTA.37560260W}.  A few LIME-olivine (Mg$\simeq$1.0)
condensates were identified \citep{2012M&PS...47..471J, 2014GeCoA.142..240F}.  Terminal particles
are also composed of other minerals, with sulfides (pyrrhotite, Fe$_{1-x}$S where $x=0-0.2$) being next in abundance
compared to olivines and pyroxenes. Other terminal particle mineral species are detailed in
\citet{2024SSRv..220...79Z} and \citet{2017M&PS...52.1612J}. These terminal particle crystals are
too large for IR emission spectroscopy to ascertain their mineralogy, as comae spectral emission
features are predominantly from submicron particles.

Submicron particles in the Stardust tracks identified as being cometary are sparsely studied compared to
terminal particles because the submicron constituents disaggregated, dispersed along the walls of
the upper bulbous parts of tracks (closer to particle entry), and intermixed with molten aerogel.
Amorphous silicates were sought in the track bulbs, but they were largely not discernible
from the aerogel \citep{2008Sci...319..447I, 2012GeCoA..87...35S, 2019M&PS...54..202I}. 
Some submicron constituents were found in regions of lesser heating and
modification: in microtracks perpendicular to the walls of the bulbous part of the tracks
\citep{2013M&PS...48.1607L}, in cracks in the aerogel nearer the terminal particles
\citep{2012LPI....43.2551N}, in the bay behind a large terminal particle whose forward-facing side
took the shock heating while hot gas extruded around its rear \citep{2019M&PS...54.1069G}, and in
the lee of a 50~\micron{}-sized terminal particle \citep{2026LPICo3160.1730I}.

The submicron cometary particulates that were studied are broadly segregated into three categories:
mineral crystals or microcrystals; equilibrated aggregates (EAs), so named by analogy to those in
chondritic anhydrous interplanetary dust particles (CA IDPs; sometimes called chondritic porous
IDPs, CP IDPs, although not all chondritic IDPs are porous); and amorphous and siliceous,
spherule-shaped materials that are GEMS-like (Glass with Embedded Metal and Sulfides).

\subsubsection{Equilibrated Aggregates}
In studies of Stardust samples, equilibrated aggregates are microcrystalline conglomerations of
primarily Mg-rich olivine (forsterite) and Mg-rich low-Ca pyroxene (enstatite), with minor diopside
and Mg-Fe-Al-Cr-spinel \citep{2013M&PS...48.1607L}, or microcrystalline conglomerations of
forsterite, enstatite, FeS or pyrrhotite, Mg-Al-Na, and Si-rich mesostasis 
\citep[glass;][]{2012LPI....43.2551N}. \citet{2005ASPC..341..657K, 2010M&PSA..73.5359K} refined the
classification into Type~1 EAs, submicrometer aggregates strictly of Mg-rich forsterite or mixed
forsterite-enstatite crystallites (50--200~nm) with equilibrium grain boundaries, and Type~2 EAs
that better match \citet{1994Sci...265..925B}'s original definition of Fe-bearing olivine and
pyroxene along with Fe-sulfides set in a silica-rich amorphous matrix. Type~1 EAs are hypothesized
to represent aggregate clusters of high-temperature nebular condensates (Mg-rich, low oxygen
fugacity) that may have undergone slight annealing. In contrast, Type~2 EAs are hypothesized 
to result from subsolidus metastable crystallization via heating (to $\sim$1000~K) of
porous, amorphous siliceous precursors, where interdiffusion is enhanced by their porous structure
\citep{2015ApJ...801L...7R}, forming Fe-olivine crystallites, FeS, remnant glass
\citep{2010M&PSA..73.5359K}, and silica \citep{2015M&PS...50.1767R}.  
Note that not all investigators use the Type~1 versus Type~2 EA terminology; some describe EAs 
as having forsteritic olivine without explicitly labeling them Type~1.

\subsubsection{Amorphous Silicates or GEMS}

GEMS are notably underdense because of high porosity at the nanoscale and non-stoichiometric (Si- or
SiO- or SiO$_{2}$-richer), with embedded Fe$_{1-x}$S (pyrrhotite) and Fe,Ni metal and often associated with
carbonaceous materials with cold-origin signatures such as D-enrichments \citep{2000Natur.404..968M}
or $^{15}$N-enrichments \citep{2018PNAS..115.6608I, 2022GeCoA.335..323B}. GEMS are three-component
assemblages in which nanometer-sized metal and sulfide inclusions are suspended in an amorphous
silicate matrix with chondritic but highly variable composition, and they retain their
characteristic nanoscale porosity over a well-defined size distribution 
\citep[0.05--0.5~\micron;][]{2021GeCoA.310..320O, 2022GeCoA.335..323B}. GEMS have amorphous silicates, but not all
amorphous silicates are GEMS in the sense of this original definition \citep{2022GeCoA.335..323B}.

Amorphous silicates are ubiquitous in CA IDPs in highly variable concentrations and with
heterogeneous compositions. In eight IDPs, 105 distinct regions of amorphous Mg-silicate glass
acting as matrix for nanophase inclusions of Fe-Ni-metal and Fe-sulfides were studied to
characterize the elemental composition of GEMS \citep{2024GeCoA.378..153S}. Most GEMS occur as
aggregates of subgrains rather than as single grains. These (new) `GEMS amorphous silicates' have
Mg:Si ratios ranging from 0.01 to 1.34, with an average of 0.43$\pm$0.03 --- well below the solar
value of 1.03. This population is dominantly Si-enriched relative to solar, so a supersolar
component is thought to be needed for CA IDPs to equate to solar (CI) composition
\citep{2024GeCoA.378..153S}. While there are differences between the earlier and more recent
definitions of GEMS \citep{2022GeCoA.335..323B, 2024GeCoA.378..153S}, the enhanced Si/Mg
compositions of GEMS are an interesting key characterization.

In terrestrial laboratory experiments,  controlled heating of GEMS produces (at 700$^\circ$C) moderately 
Fe-rich olivine and pyroxene set in a silica-rich glass --- the same mineralogy that defines Type~2 EAs. 
\citet{2005LPI....36.2391B} describe these heated-GEMS products as ``equilibrated aggregate or 
coarse-grained silicate material.'' Heating of GEMS did not produce Type~1 EAs. Both types of EAs have 
components expected from an equilibrated formation environment, but the Mg-rich Type 1 EAs require 
a lower oxygen fugacity than the Type~2 
EAs \citep{1972GeCoA..36..597G, 2016M&PS...51..843F, 2017RSPTA.37560260W}.

GEMS are sought in Stardust because, being so primitive, they are a clear signature of cometary
origins.  However, GEMS are also the most easily modified primitive material, which makes definitive
identification in the Stardust tracks even harder.  Recently, behind one of the largest (50~\micron)
terminal particles, \citet{2026LPICo3160.1730I} reported finding a submicron amorphous silicate-rich
spherule with interior Fe and Ni metal consistent with partially decomposed FeNi sulfide beads.  The
spherule also showed S-depletion at the surface, where S loss is an expected consequence of heating,
and a carbon-rich halo that might be indigenous.  It is tantalizing to consider this a GEMS, but the
evidence is not convincing. Thus, {\it bona fide} GEMS have not yet been identified in Stardust tracks
and the search continues.

\subsubsection{Microcrystals in Microtracks or Bays of Terminal Particles}
Of the four locations introduced above, the microtracks perpendicular to the bulb walls of Track
\#10 potentially offer the most direct comparison to JWST IR spectra.  They are prepared by compressing
the 3-d aerogel to 2-d, embedding in viscous epoxy or acrylic, followed by ultramicrotomy (100~nm),
which preserves the spatial association of components.  After accounting for enhanced Si from the
mixing with aerogel that occurred upon deceleration and heating, the fragment compositions comprise
25\% CI-like (solar composition), 40\% Fe and S, and 35\% crystalline silicates and Mg-bearing, Al-bearing
oxides \citep{2013M&PS...48.1607L}. The fragment-diameter distribution spans 0.01--0.12~\micron{} with
a pronounced peak at 0.04~\micron{} \citep[Figure~5]{2013M&PS...48.1607L}. At this size, extraction from
the aerogel is not feasible. They appear to have survived without compositional change; they are
polycrystalline assemblages of dominantly olivine and low-Ca pyroxene with minor Mg-Fe-Al-Cr-spinel
and diopside.  In situ studies reveal them as individual crystals or microcrystalline assemblages
with similar Mg:Fe compositions, classified as Type~I EAs following CA IDP 
nomenclature.  
The submicron olivine and low-Ca pyroxene crystals have Mg:Fe = 0.90:0.10--0.85:0.15
\citep{2009LPI....40.1785L, 2013M&PS...48.1607L}, which is in the compositional range that we
discern in the JWST data (see Section~\ref{sec:dw-dust-bestfit}).

In the same microtracks of Track \#10, the 25\% CI-like fragments are glassy patches with 10\%
cometary material and 90\% melted aerogel; their silica-rich composition, mainly Fe, Mg, and S, is
near-CI.  The CI-like fragments could be thermally modified EAs or GEMS yet do not show a loss of
sulfur \citep{2015M&PS...50.1767R}.  The preserved CI composition at this small scale argues that
this material has not been chemically differentiated by high-temperature solar-nebula processes that
would have driven off sulfur \citep[Figure~7]{2013M&PS...48.1607L}.

Terminal particle 2 of Track \#168 from aerogel cell C2098 was an 8~\micron{} aggregate with a
concentric structure: a core of 0.1~\micron{} olivine crystals co-existing with indigenous amorphous
SiO$_2$, surrounded by a 0.8~\micron{} carbon mantle, all encased in compressed aerogel
\citep{2012LPI....43.2551N}.  Other similarly studied terminal particles are likely type II
chondrule fragments.  The aerogel cracks in Track \#147, also from aerogel cell C2098, are along the
track walls.  In cracks in Track \#147 from less-heated regions 200~\micron{} from the track wall, there
are 0.5--1~\micron{} irregularly shaped polycrystalline assemblages with 0.01--0.2~\micron{} crystals of
forsterite, enstatite, diopside, and pyrrhotite, together with an amorphous Mg-Al-Na, Si-rich
mesostasis whose textures suggest it was never molten, unlike typical aerogel melts.  Three of these
polycrystalline grains were identified as like Type~I EAs \citep{2012LPI....43.2551N}.

In the bay of the large terminal FeS grain `Andromeda', 12 fine-grain-material (FGM) regions were
studied; two were ascertained to be indistinguishable from EAs and one to be GEMS-like
\citep{2019M&PS...54.1069G}.  These three represent CA IDP components found in Stardust tracks. The
micro-chondrule Febo had FGM \citep{2012M&PS...47..471J}, but further study demonstrated its
composition falls well outside the CI-compositional bounds of GEMS in CA IDPs and so this FGM was
deemed to not be GEMS \citep{2019M&PS...54..202I}.

In a separate search, micron-to-submicron crystalline silica was sought and found in 5/25 tracks
examined. The samples were 100~nm ultramicrotomed slices of samples from two aerogel cells (C2009
and one slice from C2044), embedded in epoxy or acrylic.  The presence of silica is very unusual in
extraterrestrial samples \citep{2015ApJ...801L...7R}.  The crystalline structure of the silica and
its association with adjacent minerals indicate that silica results from subsolidus metastable
crystallization, forming olivine with silica as a byproduct of long-duration annealing near 1000~K
\citep{2009ApJ...707L.174R, 2015ApJ...801L...7R}, presumably in the protoplanetary disk prior to
incorporation into the comet.

\subsection{JWST and Stardust as Complements}\label{sec-dw-dust-stardust-partB}

\begin{figure*}[ht!]
\begin{center}
\includegraphics[trim=0.07cm 0.05cm 0.07cm 0.07cm, clip, width=0.85\textwidth]{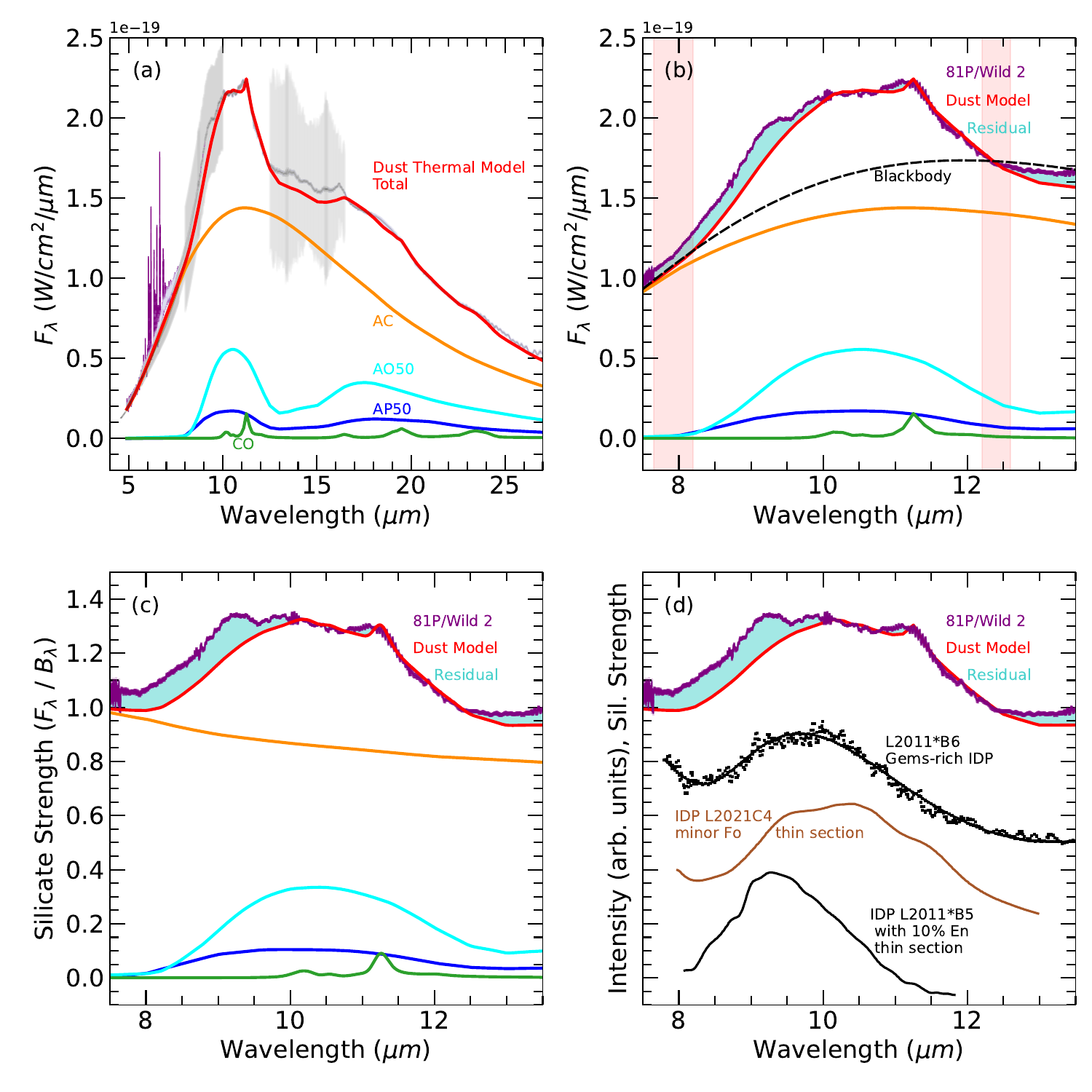}
\caption{Comet 81P/Wild~2's $10$~\micron\ silicate feature compared
to GEMS-rich IDPs. (a) MRS coma flux ({\it dimgray}) and the best-fit
dust thermal model total ({\it red}) with dust components of amorphous
carbon (AC, {\it darkorange}), amorphous olivine (AO50, {\it cyan}),
amorphous pyroxene (AP50, {\it blue}), and crystalline olivine
(CO, {\it forestgreen}). (b) A scaled blackbody ({\it black} dashed
line) fitted to the `pseudo-continuum' flux points on either side of
the $10$~\micron\ silicate feature ({\it red} shaded wavelength
regions). (c) The Silicate Strength, which is the coma flux divided by
the scaled blackbody. The dust thermal model total and its components
also are blackbody-divided to show how amorphous olivine (AO50) and
amorphous pyroxene (AP50) contribute to the Silicate Strength shape,
and to highlight how the `baseline' includes amorphous carbon and
silicates that have non-negligible opacities at
$\lambda > 12.5$~\micron. The residual of the coma minus the thermal
dust model is also shown ({\it teal} shading;
Figure~\ref{fig:PSJ-dwdust-4-20260528}). The scattered light model does not
contribute at these wavelengths. (d) The Intensity (in arbitrary
units) of IR absorption features of a GEMS-rich IDP (L2011*B6) and
GEMS-rich IDP thin sections for L2021C4 with minor forsterite (minor
Mg-crystalline olivine) and for L2011*B5 with 10\% enstatite
(Mg-crystalline orthopyroxene)
\citep[Figs.~3,~4]{1999Sci...285.1716B}. The JWST $10$~\micron\
silicate feature shape best compares to the GEMS-rich IDP thin section
with minor crystalline forsterite ({\it saddlebrown}).}
\label{fig:PSJ-dwdust-7-20260528}
\end{center}
\end{figure*}

The fine-grained crystalline silicates (submicron Stardust materials) that survived capture in
regions of lesser heating are Mg-rich olivine and low-Ca pyroxene with Mg/(Mg+Fe) = 0.90--0.85
\citep{2013M&PS...48.1607L}, which is precisely the compositional range that the JWST thermal dust
model sees in the coma of comet 81P (Figure~\ref{fig:PSJ-dwdust-7-20260528}(a,b); Section~\ref{sec:dw-dust-bestfit}).  
The thermal model also has abundant amorphous silicates that resemble the IR spectrum of a 
GEMS-rich IDP with Mg-rich crystalline olivine (Figure~\ref{fig:PSJ-dwdust-7-20260528}(c)).

Stardust contains Type~1 and Type~2 EAs. Type~1 EAs have Mg-rich silicates, consistent with the
thermal model results for 81P's coma dust emission; however, enstatite is excluded (at 95\%
confidence) from the thermal model. Some Stardust EA-like assemblages have forsterite, enstatite,
and iron-sulfides \citep[pyrrhotite;][]{2012LPI....43.2551N}, and there are nanoscale Fe-Ni-S beads in
Stardust tracks that appear to be inherited from a population of nanoscale iron-sulfides similar to
those found in cometary IDPs \citep{2012M&PS...47..594S, 2012LPI....43.1214S}. The JWST spectrum
nonetheless lacks the resonances of submicron sulfides, by comparison with spectra of FeS in an
external protoplanetary disk and with laboratory spectra of pyrrhotite 
\citep[Fe$_{1-x}$S;][]{2002Natur.417..148K}. Type~2 EAs have Fe-rich silicates, which are also clearly absent from
the JWST spectrum: submicron powders of more fayalitic olivine (Mg:Fe $\lesssim$ 0.6:0.4) have
wavelength positions of their resonances significantly different from those of the Mg-rich olivines
\citep{2003A&A...399.1101K} and so would be distinguishable if present. Stardust EAs that are
similar to Type~1 EAs are therefore mostly consistent with the thermal model. Type~2 EAs are not
found in the JWST spectrum.

The enhanced Si/Mg compositions of GEMS are an interesting key characterization because the mid-IR 
absorption spectrum of a GEMS-rich IDP (Figure~\ref{fig:PSJ-dwdust-7-20260528}(d)) has a shorter
wavelength onset for its $10$~\micron\ feature than the spectral shape of amorphous pyroxene 
(AP50; Figure~\ref{fig:PSJ-dwdust-7-20260528}(c)), computed using the optical constants for Mg:Fe=50:50 
amorphous pyroxene \citep{1995A&A...300..503D} that are widely used to model cometary IR spectra.
Additional IR spectra, and spectra that span the full range of JWST wavelengths, of GEMS or GEMS-rich 
regions of CA IDPs are needed for more detailed comparisons with IR spectra of comets.

GEMS-like materials are sought in the Stardust submicron inventory even though {\it bona fide} GEMS
have not been confirmed.  Whether remote sensing JWST spectroscopy can be used to ascertain  
the composition of amorphous siliceous materials so small and porous that they survived the Stardust capture 
intact is an outstanding question. Conversely, the Stardust samples have siliceous materials too large to impact 
the SED of JWST spectra.  JWST is thus an interesting complement to Stardust, and analysis of the remote 
sensing SED demonstrated that Mg:Fe amorphous silicates are a prominent component in the fine-grained 
submicron materials in the coma of comet 81P/Wild 2.

%% file: P81-dust-model-table1.tex

\begin{deluxetable*}{@{\extracolsep{0pt}}ccccccccccc}
\setlength{\tabcolsep}{3pt}
\tablenum{3}
\tabletypesize{\footnotesize}
\tablewidth{0pc}
\tablecaption{Comet 81P/Wild~2: Best-Fit Thermal Emission Model Parameters \label{tab:composition-modelparams}}
\tablehead{
& \\[-1.5mm]
& & & & \multicolumn{4}{c}{$N_p (\times 10^{18}$)\tablenotemark{\,a}} \\
\cline{5-8}
& & & & \colhead{Amorphous} & \colhead{Amorphous} & \colhead{Amorphous} & \colhead{Crystalline} \\[-2.5mm]
\colhead{N\sups{b}} & \colhead{M\sups{b}} & \colhead{$a_p$ \tablenotemark{$\dagger$}} & \colhead{D\sups{b}} & \colhead{Pyroxene} & \colhead{Olivine} & \colhead{Carbon} & \colhead{Olivine} & \colhead{$N_{pts}$} & \colhead{$\chi^{2}$} & \colhead{$\chi^{2}_{\nu}$} \\[-2.5mm]
& & \colhead{($\mu$m)} & & \colhead{(AP50)} & \colhead{(AO50)} & \colhead{(AC)} & \colhead{(CO)}
}
\startdata
& \\[-0.2mm]
4.3 & 25.8 & 0.7 & 2.727 & $0.5929^{+0.0205}_{-0.0206}$ & $0.9737^{+0.0114}_{-0.0114}$ & $3.0799^{+0.0030}_{-0.0030}$ & $0.5048^{+0.0061}_{-0.0061}$ & 11846 & 23756.62 & 2.01 \\[2.5mm]
\enddata
\tablenotetext{\dagger}{ \ Derived parameter.}
\tablenotetext{a}{\ Number of grains, $N_p$, at the peak ($a_p$) of the Hanner grain-size distribution \citep[HGSD,][]{1994ApJ...425..274H}.
\tablenotetext{b}{See text for parameter definitions.}}
\end{deluxetable*}


%% file: P81-dust-model-table2.tex

\begin{deluxetable*}{@{\extracolsep{0pt}}ccccccc}
\setlength{\tabcolsep}{3pt}
\tablenum{4}
\tabletypesize{\footnotesize}
\tablewidth{0pc}
\tablecaption{Comet 81P/Wild~2: Best-Fit Mass Fractions of Submicron Grains \label{tab:mod-massfractions}}
\tablehead{
&\\[-1.5mm]
\colhead{Total\tablenotemark{$\dagger$}} & \colhead{Amorphous} & \colhead{Amorphous} & \colhead{Amorphous} & \colhead{Crystalline} & \colhead{Silicate/Carbon} & \\[-2.5mm]
\colhead{Mass} & \colhead{Pyroxene} & \colhead{Olivine} & \colhead{Carbon} & \colhead{Olivine} & \colhead{Ratio} & \colhead{$f_{\rm{cryst}}$} \\[-2.5mm]
\colhead{( kg $\times 10^{7}$)} & \colhead{( $f$(ap50) $\times 10^{-1}$)} & \colhead{( $f$(ao50) $\times 10^{-1}$ )} & \colhead{( $f$(ac) $\times 10^{-1}$ )} & \colhead{( $f$(co) $\times 10^{-1}$ )} & & \colhead{( $\times 10^{-1}$) }
}
\startdata
& \\[-0.2mm]
$0.6520^{+0.0025}_{-0.0024}$ & $1.5383^{+0.0487}_{-0.0495}$ & $2.5262^{+0.0367}_{-0.0373}$ & $3.6321^{+0.0141}_{-0.0144}$ & $2.3034^{+0.0222}_{-0.0224}$ & $1.7532^{+0.0110}_{-0.0107}$ & $3.6172^{+0.0316}_{-0.0320}$ \\[2.5mm]
\enddata
\tablecomments{\, Asymmetric uncertainties indicate values constrained to a confidence level of 95\%.}
\tablenotetext{\dagger}{ \ Derived parameter for emission within a 1\farcs41 diameter aperture centered on the nucleus.}
\end{deluxetable*}

%% file: Conclusion.tex
\section{Conclusion}\label{sec:conclusion}
Our study of comet 81P/Wild 2 contextualizes the  first compositional comparison between JWST spectroscopy of a solar system object against analysis of its returned samples. Our measurements of the volatile composition of the coma found that 81P was enriched in \ce{CO2}, depleted in CO, \ce{CH4}, and \ce{CH3OH}, and consistent with average values for all other species; furthermore, we did not find evidence for significant compositional variability across perihelion passages when comparing to previous ground-based studies \citep{DelloRusso2014}. Analysis of the nucleus retrieved an effective radius consistent with that of the shape model derived from images of the Stardust flyby. Finally, thermal modeling of the dust coma demonstrated that JWST spectroscopy is highly complementary to the analysis of the Stardust returned samples, with each mission diagnostically sensitive to materials that the other was not. These results highlight potential synergies between remote sensing spectroscopy (including with JWST) and future comet rendezvous/flyby/sample return missions.

%% file: ThermalAppendix.tex
\appendix
\restartappendixnumbering 
\section{Thermal and scattered light models for three NIRSPec scaling factors}\label{appendix1}

\begin{figure}[ht!]
\begin{center}
\includegraphics[trim=0.07cm 0.05cm 0.07cm 0.07cm, clip, width=0.85\textwidth]{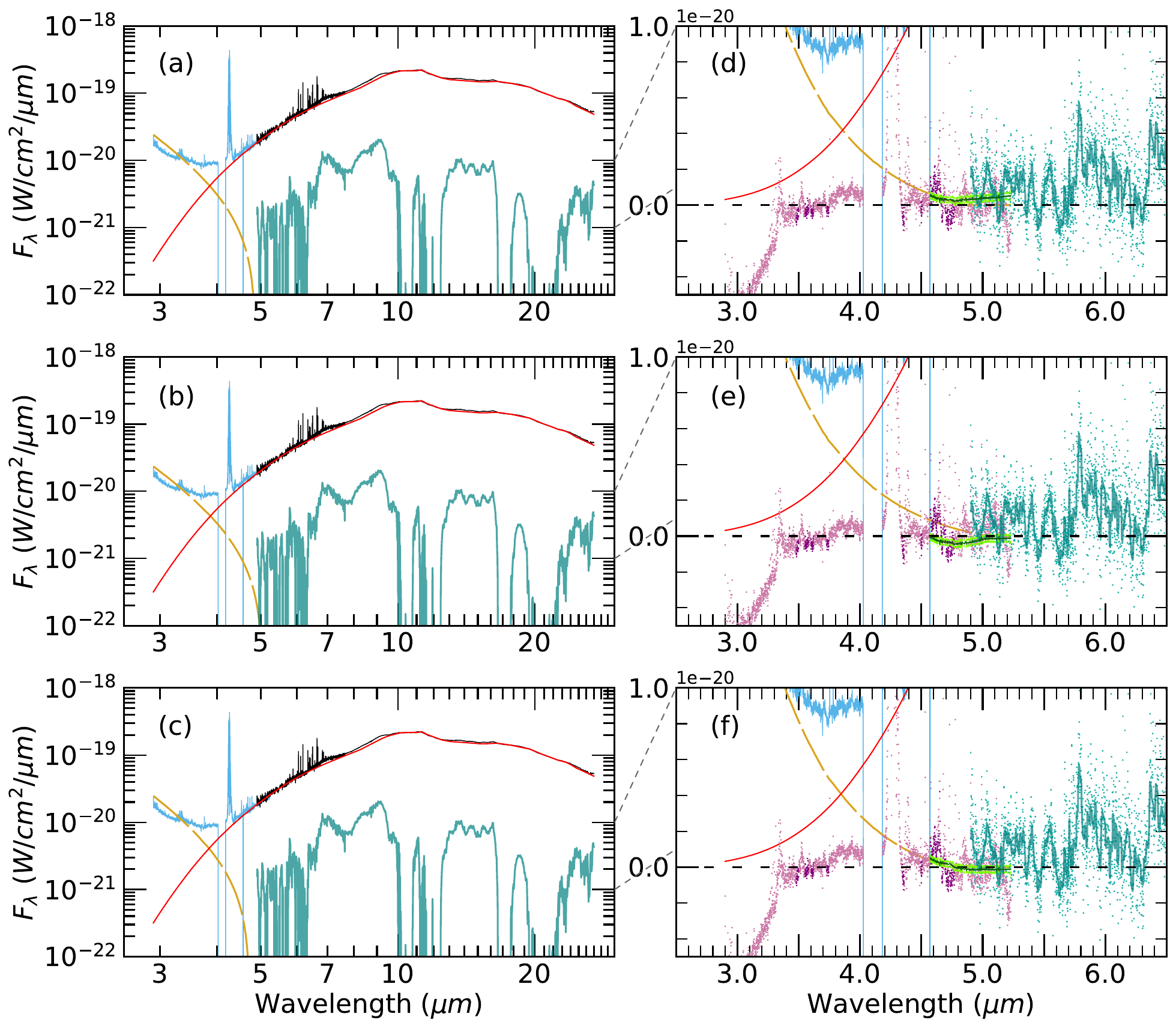}
\caption{Three sets of thermal and scattered light models for the NIRSpec coma flux scaling factors
1.042, 1.053, and 1.031. The left panels show the models plotted against the coma flux.
The right panels show the corresponding total residuals (coma flux minus the sum of the
thermal model and the scattered light model).
(a,~d) NIRSpec scaler 1.042 (0.96$\times$NEATM nucleus model).
(b,~e) NIRSpec scaler 1.053 (0.95$\times$NEATM).
(c,~f) NIRSpec scaler 1.031 (0.97$\times$NEATM).
Components shown: NIRSpec coma flux ({\it skyblue}); MRS coma flux ({\it black});
thermal dust model total ({\it red}); scattered light model ({\it olive}, long dashed);
MRS total residual, median filtered ({\it teal}); NIRSpec total residual
({\it mediumvioletred}); NIRSpec data points used in fitting the scattered light model
({\it purple}); 4.57--5.2~\micron\ PSG baseline with enhanced uncertainties
({\it darkgreen}, {\it chartreuse} error bars).
\label{fig:PSJ-dwdust-4-EXTRAS-20260530}}
\end{center}
\end{figure}

Figure~\ref{fig:PSJ-dwdust-4-EXTRAS-20260530} shows the thermal and scattered light models for the
three NIRSpec coma flux scaling factors 1.042, 1.053, and 1.031 plotted against the coma flux,
together with their total residuals. For the NIRSpec coma scale factor of 1.053, the shallower
scattered light slope, when extrapolated to MRS wavelengths, yields a slight over-subtraction
at 4.5--5.0~\micron\ in the overlap region between the two instruments. For 1.031, in the
overlap region the scattered light fit runs through the lower error bars of the
4.57--5.2~\micron\ PSG baseline with enhanced uncertainties, but its steeper scattered light
slope yields a slight over-subtraction at $\lambda > 5.0$~\micron. The 1.042 scaler yields
the best total residual in the overlap region. The differences among the total residuals for
the three scaling factors remain small compared to the MRS instrumental uncertainties adopted for
the residuals (see Section~\ref{subsubsection:dust-model}).

